\documentclass[twocolumn]{aastex631}

\newcommand{\angstrom}{\textup{\AA}}
\usepackage{graphicx}
\usepackage{amsmath, makecell, booktabs, nicematrix, enumitem}
\usepackage{threeparttable}

\shorttitle{LyCAN continuum predictions}
\shortauthors{Turner et al.}

\begin{document}

\title{New measurements of the Lyman-$\alpha$ forest continuum and effective optical depth with LyCAN and DESI Y1 data}

\correspondingauthor{Wynne Turner}
\email{turner.1839@osu.edu}

\author[0009-0008-3418-5599]{Wynne~Turner}
\affiliation{Department of Astronomy, The Ohio State University, 140 West 18th Avenue, Columbus, OH 43210, USA}
\affiliation{Center for Cosmology \& AstroParticle Physics, The Ohio State University, 191 West Woodruff Avenue, Columbus, OH 43210, USA}

\author[0000-0002-4279-4182]{Paul~Martini}
\affiliation{Department of Astronomy, The Ohio State University, 140 West 18th Avenue, Columbus, OH 43210, USA}
\affiliation{Center for Cosmology \& AstroParticle Physics, The Ohio State University, 191 West Woodruff Avenue, Columbus, OH 43210, USA}
\affiliation{Department of Physics, The Ohio State University, 191 West Woodruff Avenue, Columbus, OH 43210, USA}

\author[0000-0001-7336-8912]{Na\textsc{\.{\.i}}m~Göksel Karaçaylı}
\affiliation{Department of Astronomy, The Ohio State University, 140 West 18th Avenue, Columbus, OH 43210, USA}
\affiliation{Center for Cosmology \& AstroParticle Physics, The Ohio State University, 191 West Woodruff Avenue, Columbus, OH 43210, USA}
\affiliation{Department of Physics, The Ohio State University, 191 West Woodruff Avenue, Columbus, OH 43210, USA}

%alphabetical

\author{J.~Aguilar}
\affiliation{Lawrence Berkeley National Laboratory, 1 Cyclotron Road, Berkeley, CA 94720, USA}

\author[0000-0001-6098-7247]{S.~Ahlen}
\affiliation{Physics Department, Boston University, 590 Commonwealth Avenue, Boston, MA 02215, USA}

\author{D.~Brooks}
\affiliation{Department of Physics \& Astronomy, University College London, Gower Street, London, WC1E 6BT, UK}

\author{T.~Claybaugh}
\affiliation{Lawrence Berkeley National Laboratory, 1 Cyclotron Road, Berkeley, CA 94720, USA}

\author[0000-0002-1769-1640]{A.~de la Macorra}
\affiliation{Instituto de F\'{\i}sica, Universidad Nacional Aut\'{o}noma de M\'{e}xico,  Cd. de M\'{e}xico  C.P. 04510,  M\'{e}xico}

\author[0000-0002-4928-4003]{A.~Dey}
\affiliation{NSF NOIRLab, 950 N. Cherry Ave., Tucson, AZ 85719, USA}

\author{P.~Doel}
\affiliation{Department of Physics \& Astronomy, University College London, Gower Street, London, WC1E 6BT, UK}

\author[0000-0003-2371-3356]{K.~Fanning}
\affiliation{Kavli Institute for Particle Astrophysics and Cosmology, Stanford University, Menlo Park, CA 94305, USA}
\affiliation{SLAC National Accelerator Laboratory, Menlo Park, CA 94305, USA}

\author[0000-0002-2890-3725]{J~.E.~Forero-Romero}
\affiliation{Departamento de F\'isica, Universidad de los Andes, Cra. 1 No. 18A-10, Edificio Ip, CP 111711, Bogot\'a, Colombia}
\affiliation{Observatorio Astron\'omico, Universidad de los Andes, Cra. 1 No. 18A-10, Edificio H, CP 111711 Bogot\'a, Colombia}

\author[0000-0003-3142-233X]{S.~Gontcho A Gontcho}
\affiliation{Lawrence Berkeley National Laboratory, 1 Cyclotron Road, Berkeley, CA 94720, USA}

\author[0000-0003-4089-6924]{A.~X.~Gonzalez-Morales}
\affiliation{Consejo Nacional de Ciencia y Tecnolog\'{\i}a, Av. Insurgentes Sur 1582. Colonia Cr\'{e}dito Constructor, Del. Benito Ju\'{a}rez C.P. 03940, M\'{e}xico D.F. M\'{e}xico}
\affiliation{Departamento de F\'{i}sica, Universidad de Guanajuato - DCI, C.P. 37150, Leon, Guanajuato, M\'{e}xico}

\author{G.~Gutierrez}
\affiliation{Fermi National Accelerator Laboratory, PO Box 500, Batavia, IL 60510, USA}

\author[0000-0001-9822-6793]{J.~Guy}
\affiliation{Lawrence Berkeley National Laboratory, 1 Cyclotron Road, Berkeley, CA 94720, USA}

\author[0000-0002-9136-9609]{H.~K.~Herrera-Alcantar}
\affiliation{Departamento de F\'{i}sica, Universidad de Guanajuato - DCI, C.P. 37150, Leon, Guanajuato, M\'{e}xico}

\author{K.~Honscheid}
\affiliation{Center for Cosmology \& AstroParticle Physics, The Ohio State University, 191 West Woodruff Avenue, Columbus, OH 43210, USA}
\affiliation{Department of Physics, The Ohio State University, 191 West Woodruff Avenue, Columbus, OH 43210, USA}

\author{S.~Juneau}
\affiliation{NSF NOIRLab, 950 N. Cherry Ave., Tucson, AZ 85719, USA}

\author[0000-0003-3510-7134]{T.~Kisner}
\affiliation{Lawrence Berkeley National Laboratory, 1 Cyclotron Road, Berkeley, CA 94720, USA}

\author[0000-0001-6356-7424]{A.~Kremin}
\affiliation{Lawrence Berkeley National Laboratory, 1 Cyclotron Road, Berkeley, CA 94720, USA}

\author{A.~Lambert}
\affiliation{Lawrence Berkeley National Laboratory, 1 Cyclotron Road, Berkeley, CA 94720, USA}

\author[0000-0003-1838-8528]{M.~Landriau}
\affiliation{Lawrence Berkeley National Laboratory, 1 Cyclotron Road, Berkeley, CA 94720, USA}

\author[0000-0001-7178-8868]{L.~Le~Guillou}
\affiliation{Sorbonne Universit\'{e}, CNRS/IN2P3, Laboratoire de Physique Nucl\'{e}aire et de Hautes Energies (LPNHE), FR-75005 Paris, France}

\author[0000-0002-1125-7384]{A.~Meisner}
\affiliation{NSF NOIRLab, 950 N. Cherry Ave., Tucson, AZ 85719, USA}

\author{R.~Miquel}
\affiliation{Instituci\'{o} Catalana de Recerca i Estudis Avan\c{c}ats, Passeig de Llu\'{\i}s Companys, 23, 08010 Barcelona, Spain}
\affiliation{Institut de F\'{i}sica d'Altes Energies (IFAE), The Barcelona Institute of Science and Technology, Campus UAB, 08193 Bellaterra Barcelona, Spain}

\author[0000-0002-2733-4559]{J.~Moustakas}
\affiliation{Department of Physics and Astronomy, Siena College, 515 Loudon Road, Loudonville, NY 12211, USA}

\author{E.~Mueller}
\affiliation{Department of Physics and Astronomy, University of Sussex, Brighton BN1 9QH, UK}

\author{A.~Mu\~noz-Guti\'{e}rrez}
\affiliation{Instituto de F\'{\i}sica, Universidad Nacional Aut\'{o}noma de M\'{e}xico,  Cd. de M\'{e}xico  C.P. 04510,  M\'{e}xico}

\author{A.~D.~Myers}
\affiliation{Department of Physics \& Astronomy, University  of Wyoming, 1000 E. University, Dept.~3905, Laramie, WY 82071, USA}

\author[0000-0001-6590-8122]{J.~Nie}
\affiliation{National Astronomical Observatories, Chinese Academy of Sciences, A20 Datun Rd., Chaoyang District, Beijing, 100012, P.R. China}

\author[0000-0002-1544-8946]{G.~Niz}
\affiliation{Departamento de F\'{i}sica, Universidad de Guanajuato - DCI, C.P. 37150, Leon, Guanajuato, M\'{e}xico}
\affiliation{Instituto Avanzado de Cosmolog\'{\i}a A.~C., San Marcos 11 - Atenas 202. Magdalena Contreras, 10720. Ciudad de M\'{e}xico, M\'{e}xico}

\author{C.~Poppett}
\affiliation{Lawrence Berkeley National Laboratory, 1 Cyclotron Road, Berkeley, CA 94720, USA}
\affiliation{Space Sciences Laboratory, University of California, Berkeley, 7 Gauss Way, Berkeley, CA  94720, USA}

\author[0000-0001-7145-8674]{F.~Prada}
\affiliation{Instituto de Astrof\'{i}sica de Andaluc\'{i}a (CSIC), Glorieta de la Astronom\'{i}a, s/n, E-18008 Granada, Spain}

\author[0000-0001-5589-7116]{M.~Rezaie}
\affiliation{Department of Physics, Kansas State University, 116 Cardwell Hall, Manhattan, KS 66506, USA}

\author{G.~Rossi}
\affiliation{Department of Physics and Astronomy, Sejong University, Seoul, 143-747, Korea}

\author[0000-0002-9646-8198]{E.~Sanchez}
\affiliation{CIEMAT, Avenida Complutense 40, E-28040 Madrid, Spain}

\author[0000-0002-3569-7421]{E.~F.~Schlafly}
\affiliation{Space Telescope Science Institute, 3700 San Martin Drive, Baltimore, MD 21218, USA}

\author{D.~Schlegel}
\affiliation{Lawrence Berkeley National Laboratory, 1 Cyclotron Road, Berkeley, CA 94720, USA}

\author{M.~Schubnell}
\affiliation{Department of Physics, University of Michigan, Ann Arbor, MI 48109, USA}

\author[0000-0002-6588-3508]{H.~Seo}
\affiliation{Department of Physics \& Astronomy, Ohio University, Athens, OH 45701, USA}

\author{D.~Sprayberry}
\affiliation{NSF NOIRLab, 950 N. Cherry Ave., Tucson, AZ 85719, USA}

\author[0000-0003-1704-0781]{G.~Tarl\'{e}}
\affiliation{Department of Physics, University of Michigan, Ann Arbor, MI 48109, USA}

\author{B.~A.~Weaver}
\affiliation{NSF NOIRLab, 950 N. Cherry Ave., Tucson, AZ 85719, USA}

\author[0000-0002-6684-3997]{H.~Zou}
\affiliation{National Astronomical Observatories, Chinese Academy of Sciences, A20 Datun Rd., Chaoyang District, Beijing, 100012, P.R. China}

\begin{abstract}
We present the Lyman-$\alpha$ Continuum Analysis Network (LyCAN), a Convolutional Neural Network that predicts the unabsorbed quasar continuum within the rest-frame wavelength range of $1040-1600\,\angstrom$ based on the red side of the Lyman-$\alpha$ emission line ($1216-1600\,\angstrom$). We developed synthetic spectra based on a Gaussian Mixture Model representation of Nonnegative Matrix Factorization (NMF) coefficients. These coefficients were derived from high-resolution, low-redshift ($z<0.2$) Hubble Space Telescope/Cosmic Origins Spectrograph quasar spectra. We supplemented this COS-based synthetic sample with an equal number of DESI Year 5 mock spectra. LyCAN performs extremely well on testing sets, achieving a median error in the forest region of 1.5\% on the DESI mock sample, 2.0\% on the COS-based synthetic sample, and 4.1\% on the original COS spectra. LyCAN outperforms Principal Component Analysis (PCA)- and NMF-based prediction methods using the same training set by 40\% or more. We predict the intrinsic continua of 83,635 DESI Year 1 spectra in the redshift range of $2.1 \leq z \leq 4.2$ and perform an absolute measurement of the evolution of the effective optical depth. This is the largest sample employed to measure the optical depth evolution to date. We fit a power-law of the form $\tau(z) = \tau_0 (1+z)^\gamma$ to our measurements and find $\tau_0 = (2.46 \pm 0.14)\times10^{-3}$ and $\gamma = 3.62 \pm 0.04$. Our results show particular agreement with high-resolution, ground-based observations around $z = 2$, indicating that LyCAN is able to predict the quasar continuum in the forest region with only spectral information outside the forest.
\end{abstract}

\keywords{Convolutional neural networks (1938) --- Cosmology (343) --- Dark energy (351) --- Intergalactic medium (813) -- Large-scale structure of the universe (902) -- Lyman alpha forest (980)}

\section{Introduction} \label{sec:intro}

Observations indicate that the Universe is expanding at an accelerated rate. Measurements of Type 1a supernovae provided the first evidence for this accelerated expansion \citep{riess_1998_supernovae,perlmutter_1999_supernovae}, the cause of which is unknown and called dark energy. Since then, multiple cosmological surveys have contributed to the study of the accelerating expansion, including observations of the Cosmic Microwave Background \citep[e.g.][]{planck_2016} and measurements of the Baryon Acoustic Oscillation (BAO) standard ruler with galaxy and quasar spectra \citep[e.g.][]{2df,sdss_bao_2005}. \cite{weinberg_review_2013} provided a comprehensive review of observational probes of cosmic acceleration. 

Quasars are among the most luminous objects in the Universe, acting as cosmic flashlights that illuminate the matter-density distribution. The Ly$\alpha$ forest, a collection of absorption lines detected blueward of the Ly$\alpha$ emission line in the spectra of distant quasars, captures this matter-density distribution along the line of sight. The first to measure the BAO scale with the Lyman-alpha (Ly$\alpha$) forest was the Baryon Oscillation Spectroscopic Survey \citep[BOSS;][]{dawson_boss_2013} program within the third stage of the Sloan Digital Sky Survey \citep[SDSS-III;][]{eisenstein_sdss3_2011} \citep[e.g.][]{busca_2013_lya_boss,slosar_2013_lya_boss,fontribera_2014_lyaqso_boss}. The successor to BOSS, extended-BOSS \citep[eBOSS;][]{dawson_eboss_2016} as a part of SDSS-IV \citep{blanton_sdss4_2017}, performed the same measurements on a sample of $\sim4$ million galaxies and quasars \citep[e.g.][]{ahumada_eboss_2020,bourboux_completed_2020}.

While BOSS and eBOSS made great strides in constraining cosmological parameters, the nature of dark energy remains unknown. The Dark Energy Spectroscopic Instrument (DESI), the largest spectroscopic survey to date, aims to achieve the tightest constraints on the dark energy equation of state by measuring spectra of $\sim40$ million galaxies and quasars over a five-year period \citep[][]{desi_collaboration_desi_2016, desi_instrument_2016}. Roughly one million DESI targets are devoted to the high-redshift ($z \geq 2.1$) quasars used to constrain cosmological parameters with the Ly$\alpha$ forest. Each forest absorption feature provides a measurement of the matter-density distribution of the intergalactic medium (IGM) along the line of sight to the quasar. At the DESI resolution, this can correspond to well over a hundred independent absorption measurements per quasar if the entire forest is visible, whereas a quasar used as a discrete tracer only provides one measurement of the matter-density field.

The Ly$\alpha$ forest can also be used to probe properties of the IGM, such as its temperature and ionization state \citep[e.g.][]{schaye_2000,zaldarriaga_2001,mcdonald_2001,bolton_2005,meiksin_2009,becker_detection_2011,becker_2015,walther_2019,khaire_2019,gaikwad_2021}. Measurements of the effective optical depth of the forest $\tau_{\rm eff}$ are one way to study these properties, and the scatter of the optical depth also reveals information about the intensity of the ionizing background radiation produced by active galactic nuclei and star-forming galaxies \citep[e.g.][]{becker_2015,kulkarni_2019}. Other works have also studied the escape fraction of ionizing photons \citep[e.g.][]{khaire_2016} and the evolution of the $UV$ background, including with the development of empirically-constrained models \citep[e.g.][]{khaire_srinand_2019,puchwein_2019,faucher_giguere_2020}. \cite{mcquinn_review_2016} provided a comprehensive review on the evolution of the IGM. While the Ly$\alpha$ forest is readily accessible at $z>2$ with ground based observations, the optical depth is not straightforward to measure because it is difficult to determine the unabsorbed quasar continuum due to the forest absorption.

In order to exploit the Ly$\alpha$ forest for cosmological analysis and studies of the IGM, accurate knowledge of the quasar continuum is required. This becomes increasingly difficult at higher redshifts, since the increase in the neutral fraction and density of the high-redshift Universe prohibits direct measurements of the true continuum in the forest region. In the BOSS/eBOSS and current DESI Ly$\alpha$ analysis pipeline, \texttt{picca}\footnote{\url{https://github.com/igmhub/picca/}}, the flux transmission field $\delta_q(\lambda)$ is measured in the Ly$\alpha$ forest region as a function of observed wavelength $\lambda$ according to
\begin{equation}\label{deltas}
    \delta_q(\lambda) = \frac{f_q(\lambda)}{\Bar{F}(\lambda)C_q(\lambda)} - 1
\end{equation}
where $f_q(\lambda)$ is the observed flux, and the mean expected flux $\Bar{F}(\lambda)C_q(\lambda)$ is the product of the mean transmission $\Bar{F}(\lambda)$ and the unabsorbed continuum $C_q(\lambda)$ \citep{bourboux_completed_2020}. These deltas $\delta_q(\lambda)$ are measured for each Ly$\alpha$ forest pixel and then auto-correlated with each other or cross-correlated with quasars to achieve a 3D correlation function. Lacking knowledge of the true continuum, a two-parameter fit of the mean continuum $\Bar{F}(\lambda)C_q(\lambda)$ in the forest region is performed on the observed flux. However, this introduces spurious correlations that complicate the analysis since the fitting process biases the mean of each delta toward zero for each line of sight. This biasing also results in the loss of information on large scales and projects out large-scale modes, which impacts the power spectrum in one dimension \citep[e.g.][]{karacayli_optimal_2020,karacayli_optimal_2022,karacayli_optimal_2024,ravoux_dark_2023} and in three dimensions \citep[e.g.][]{karim_2023_p3d,roger_p3d_2024}. The standard approach is now to use a distortion matrix to accurately model and account for these spurious correlations. \cite{bautista_2017} provide a detailed description of the distortion matrix.

There have been a number of attempts to measure the unabsorbed Ly$\alpha$ forest continuum. Some previous works have traced the transmission peaks in the Ly$\alpha$ forest region in high-resolution spectra to estimate the true continuum \citep[e.g.][]{rauch_opacity_1997,schaye_metallicity_2003}, but this is likely inaccurate for lower resolution spectra and at higher redshifts where the transmission peaks do not represent the unabsorbed continuum. Others have used composites of medium-resolution spectra to measure the continuum \citep[e.g.][]{becker_refined_2013}, but this is also not a direct measurement of the continuum at high redshifts.

Because a correlation exists between the continuum shape including broad emission lines on the red and blue sides of the Ly$\alpha$ emission line \citep[e.g.][]{paris_principal_2011,greig_2017_emissionlines}, other works have attempted to predict the true continuum using only longer wavelength information (redward of the Ly$\alpha$ emission line) with Principal Component Analysis \citep[PCA; e.g.][]{suzuki_predicting_2005,paris_principal_2011,lee_mean-flux_2012,davies_predicting_2018}. \cite{lee_mean-flux_2012} also corrected their continuum predictions using external constraints on the Ly$\alpha$ forest mean flux, making their method dependent on the Ly$\alpha$ forest region. In general, dimensionality reduction techniques like PCA or Nonnegative Matrix Factorization (NMF) are limited by their ability to learn linear representations of the data, while neural networks are able to learn more complex, nonlinear representations.  We will show that PCA- and NMF-based prediction models do not work as well as a deep learning approach that has the ability to learn nonlinear relationships.

Several recent efforts have illustrated the potential of deep learning approaches. Some have used such approaches to infer properties of the IGM \citep[e.g.][]{dl_huang_2021,dl_nayak_2023,dl_nasir_2024} or fit the absorption lines in the forest with Voigt profiles \citep{jalan_2024}. Additionally, deep learning can be used to predict the unabsorbed continuum in the forest region. \cite{liu_deep_2021} developed a feedforward neural network called iQNet designed to predict the quasar continuum in the Ly$\alpha$ forest region using only information on the red side. However, their training set was entirely based on COS quasars and the resulting model is likely not optimized for DESI quasars, as for example the DESI quasars are generally much more luminous. Another effort from \cite{sun_quasar_2023} modeled the quasar continuum using an unsupervised probabilistic model. This model achieves promising results, although is not independent of the observed flux in the Ly$\alpha$ forest region. The dependence on flux in the forest region also does not enable an independent measurement of the mean absorption. 

In this work, we present a Convolutional Neural Network (CNN) to predict the Ly$\alpha$ forest continuum that we call LyCAN, the Ly$\alpha$ Continuum Analysis Network. We used a Gaussian Mixture Model (GMM) representation of the NMF coefficients of low-$z$ $(z<0.2)$ Hubble Space Telescope (HST)/Cosmic Origins Spectrograph (COS) spectra to develop a sample of 40,000 synthetic spectra. We altered the emission line strengths of half of these spectra so that they are more representative of high-luminosity quasars. We trained our neural network on a random subset of this sample of 40,000 COS-based synthetic spectra combined with an equal number of DESI Year 5 (Y5) mock spectra \citep[e.g.][]{farr_lyacolore_2020,herrera-alcantar_synthetic_2023} and used the resulting network to predict the Ly$\alpha$ forest continua for the training and testing samples, a larger sample of DESI Y5 mock spectra, and the real quasar spectra from DESI Year 1 (Y1).

This paper is organized as follows: in Section~\ref{sec:data} we introduce the data used in this study, including the low-$z$ HST/COS spectra, DESI mocks, and DESI Y1 spectra. In Section~\ref{sec:synth} we detail the generation of our synthetic spectra. These synthetic spectra were generated from archival COS observations of low-redshift quasars, which we augmented with mock spectra generated for the DESI Lyman-$\alpha$ forest analysis. We discuss the creation and performance of our neural network in Section~\ref{sec:lycan}, and compare its performance with PCA- and NMF-based prediction models. These PCA and NMF models use the same training set as LyCAN, and are distinct from the NMF representation that is used to help generate synthetic spectra. In Section~\ref{sec:tau}, we use the LyCAN continuum predictions to calculate the evolution of the effective optical depth on DESI Y1 quasars and compare these results to previous works in the literature. Finally, we discuss our conclusions and planned future work in Section~\ref{sec:summary}.

\section{Data}\label{sec:data}

We used several datasets for several purposes. The first purpose is the generation and assembly of a synthetic dataset for training, testing, and validation of LyCAN. This dataset includes archival observations from HST/COS. We describe this sample in Section~\ref{sec:cos}, along with our processing steps to prepare these data to create the training set. DESI observations are important to both ensure our synthetic data are representative of DESI data, and for our measurement of the mean optical depth. We describe DESI observations in Section~\ref{sec:desidata}. Finally, we supplemented the COS sample with DESI mocks to construct our synthetic dataset, as well as used synthetic mock observations to characterize uncertainties in our fit to the mean optical depth evolution. We describe these DESI mocks in Section~\ref{sec:desimocks}.

\subsection{HST/COS}
\label{sec:cos}

We used low-redshift ($z < 0.2$) quasars observed by the COS instrument on the HST to help construct our training, testing, and validation samples. Low-redshift quasars are very minimally affected by Ly$\alpha$ forest absorption, and they therefore provide the best measurement of the true continuum in that region.

We retrieved spectra from the Hubble Spectroscopic Legacy Archive \citep[HSLA;][]{peeples_hubble_2017}, and then selected the subset with rest-frame wavelength coverage from $1070-1600\,\angstrom$ and a median signal to noise ratio (SNR) of at least five per resolution element. We chose this minimum rest wavelength requirement to maximize our sample size, since very few quasars in the HSLA fully span the entire Ly$\alpha$ forest region down to 1040\,\angstrom.~Furthermore, some spectra suffer from a significant number of missing data pixels in this more narrow forest region. We removed QSOs that were missing more than 30\% of the native pixels in the minimum forest region ($1070-1200\,\angstrom$), and then manually removed suspected Broad Absorption Line (BAL) quasars. This resulted in a sample of 48 QSOs. Finally, we transformed each quasar spectrum to units of relative flux by normalizing by the mean flux in the 2\,\angstrom\, interval centered around $1280\,\angstrom$.

\begin{figure}
    \centering
	\includegraphics[width=\columnwidth]{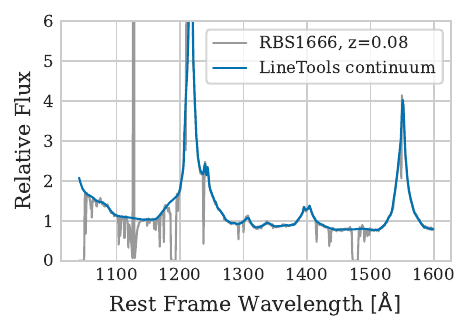}
    \caption{HST/COS quasar spectrum and the smooth representation from LineTools. The original COS spectrum is in gray and the continuum fit is in blue. The continumm includes the extrapolation to 1040\,\angstrom~(see~\S\ref{sec:preprocess}).}
    \label{fig:LT_extrap}
\end{figure}

\begin{figure}
    \centering
    \includegraphics[width=\columnwidth]{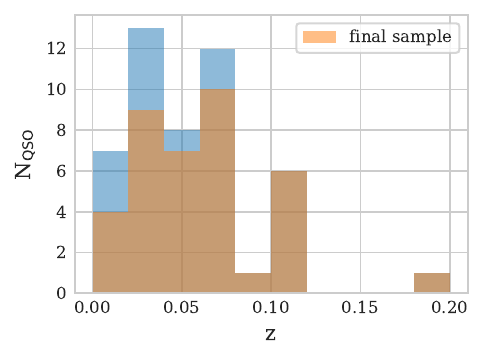}
    \caption{Redshift distribution of the HST/COS objects. The blue distribution is the original sample of 48 objects, while the orange distribution shows the final sample of 38 objects after outlier removal (see~\S\ref{sec:preprocess}).}
    \label{fig:cos_zs}
\end{figure}

Following the work of \citet{liu_deep_2021}, we used the LineTools API\footnote{\url{https://linetools.readthedocs.io}} to interactively fit a smooth continuum model to each of these quasar spectra. This included interpolating over geocoronal Ly$\alpha$ emission and Galactic absorption features. Some spectra were missing data that extend to  $1040\,\angstrom$. In these cases we extrapolated the LineTools continua to $1040\,\angstrom$ using the eBOSS DR14 \citep{blomqvist_baryon_2019,de_sainte_agathe_baryon_2019} mean quasar continuum. We also corrected each quasar for extinction due to interstellar dust. We retrieved $\rm E(B-V)$ values for each of our COS objects from \citet{schlafly_measuring_2011} and de-reddened each spectrum with the \citet{gordon_one_2023} extinction model assuming $\rm R_V=3.1$. We then eliminated 10 COS objects that appeared least similar to DESI-like quasars, which yielded a final sample of 38 COS quasars. We detail the continuum extrapolation and outlier removal procedures in Section~\ref{sec:preprocess}. Figure~\ref{fig:LT_extrap} shows an example of a LineTools continuum fit and its continuum extrapolation. The redshift distribution of the initial 48 COS objects and the final 38 objects after outlier removal is shown in Figure~\ref{fig:cos_zs}. In Section~\ref{sec:preprocess} we describe how we used this sample of COS objects to generate 40,000 synthetic spectra.

\subsection{DESI}\label{sec:desidata}

DESI was installed at the Mayall 4-m telescope at Kitt Peak National Observatory \citep{desi_collaboration_overview_2022} and is in the midst of the largest spectroscopic survey of the Universe. The aim of DESI is to measure spectra for $\sim40$ million galaxies and quasars from the local Universe to $z>3.5$ \citep{desi_collaboration_desi_2016} and study the nature of dark energy. The main cosmological tools are the BAO standard ruler to measure distances and constrain cosmological parameters, and redshift-space distortions to measure the growth of structure and test possible modifications to general relativity. DESI has 10 spectrographs and 5000 fibers that may be positioned across a 3 degree diameter field of view, which allows for efficient observations of 5000 targets simultaneously. The main DESI survey began in May 2021 upon the completion of commissioning and survey validation \citep{desi_collaboration_validation_2023}. An early data release included data from the survey validation period and the first two months of the main survey \citep{desi_collaboration_early_2023,ramirez_lya_edr}. The DESI collaboration recently completed a series of papers that describe cosmological results from the first year of data collection \citep{desi_y1_3,desi_y1_4,desi_y1_6}. Additional key papers \citep{desi_y1_2,desi_y1_5,desi_y1_7,desi_y1_8} based on Data Release 1 \citep[DR1;][]{desi_y1_dr1} are forthcoming.

The purpose of LyCAN is to predict the Ly$\alpha$ forest continuum for DESI quasars. The DESI Y1 quasar catalog contains $\sim450,000$ Ly$\alpha$ forest ($z\geq2.1$) quasars. In this paper, we demonstrate the performance of LyCAN with a subset of these quasars where the rest wavelength coverage spans at least $1160-1600\,\angstrom$. We also eliminated spectra with redshift warnings and those with Broad Absorption Lines (BALs) and/or Damped Ly$\alpha$ Absorbers (DLAs). A total of 250,955 quasars satisfy these conditions. We resampled these spectra onto a rest wavelength grid from $1040-1600\,\angstrom$ with a dispersion of $0.8\,\angstrom$ per pixel and normalized by the 40 pixels surrounding $1280\,\angstrom$. We corrected each quasar for extinction due to interstellar dust as described in Section~\ref{sec:cos}.

We used a subset of the DESI Y1 quasars with median $r-$band SNR $\geq 5$ to measure the evolution of the Ly$\alpha$ optical depth. For this subset we did not eliminate quasars with known DLA systems because we later corrected our measurements for optically thick absorbers (see~\S\ref{sec:taucorrs}). The other selection criteria were unchanged. This sample contains 83,635 quasars, which is the largest sample to date used for measurements of the effective optical depth.

\subsection{DESI Mocks}\label{sec:desimocks}

We supplemented our synthetic dataset for training, testing, and validation with DESI mock spectra that were constructed for the main cosmological analysis \citep[e.g.][]{farr_lyacolore_2020,herrera-alcantar_synthetic_2023}. The DESI mock spectra were generated with a software suite called \texttt{quickquasars}\footnote{\url{https://github.com/desihub/desisim/blob/main/py/desisim/scripts/quickquasars.py}}. The quasar continua used by \texttt{quickquasars} are generated with the \texttt{simqso}\footnote{\url{https://github.com/imcgreer/simqso}} library, which models the continuum as a broken power-law and is tuned to better reproduce the mean continuum observed in eBOSS DR16 \citep{bourboux_completed_2020} data via a Principal Component Analysis (PCA). These unabsorbed continua are multiplied by transmission files generated with the \texttt{LyaCoLoRe} program \citep{farr_lyacolore_2020}. \texttt{quickquasars} includes instrumental effects and astrophysical contaminants to produce ``fully contaminated'' mocks that include BALs, DLAs, high column density systems, and metal lines. 

The final sample of mock spectra are representative of the DESI magnitude and redshift distribution, including with realistic noise and instrumental effects. For more details about the synthesis of the DESI mocks, see \citet{herrera-alcantar_synthetic_2023}. For this work, we used the ``fully contaminated'' Year 5 (Y5) mock spectra, since these best represent the eventual quality of DESI data and already span the noise properties of Y1. The main distinction is that the Ly$\alpha$ quasars will receive multiple observations over the course of the survey, and so the Y1 quasars will eventually have somewhat higher mean SNR. We used a realization of the mocks that contains the true continua in the forest region, in addition to the simulated (observed) forest. We conducted our analysis with a random sample of 40,000 mock spectra that exclude BALs and DLAs. These spectra fully cover the rest wavelength range of $1040-1600\,\angstrom$. This sample of mock spectra spans from $2.43 < z < 2.72$.

We used a different realization of Y5 mocks with median $r-$band SNR $\geq 5$ to compute the evolution of the effective optical depth with both LyCAN continuum predictions and the truth continua. This sample consists of 311,833 mock quasars with $2.1 < z < 3.8$, and includes DLAs but not BALs.  In Section~\ref{sec:tau} we show these results and discuss how we used them to calibrate our systematic uncertainties.

\section{COS-based Synthetic Spectra}\label{sec:synth}

We constructed a sample of 40,000 synthetic spectra from our base sample of archival COS observations (see \S\ref{sec:cos}) to produce a sufficient sample to train our neural network. This process of data augmentation is a standard practice for training neural networks to increase the size of the training set, such as to help teach the network to ignore certain features like noise. We combined this COS-based synthetic sample with DESI mock spectra to construct our training, testing, and validation data sets. In Section~\ref{sec:preprocess} we detail the preprocessing steps required to generate our synthetic spectra and in Section~\ref{sec:synthspec} we describe the methods we used to create the set of 40,000 synthetic spectra.

\subsection{Preprocessing}\label{sec:preprocess}
We require full coverage of the Ly$\alpha$ forest region for training of the neural network. It is therefore important that the basis of our sample of synthetic spectra covers the full forest; however, many of the COS spectra are missing data on the blue end of the forest region. We extended these spectra to cover the full rest-frame wavelength range of $1040-1600\,\angstrom$ with the mean continuum measured by \texttt{picca} in eBOSS DR14 \citep{blomqvist_baryon_2019,de_sainte_agathe_baryon_2019}. For each COS spectrum with missing data on the blue end of the Ly$\alpha$ forest region, we performed this extrapolation as follows: we first normalized the eBOSS mean continuum to the COS LineTools continuum over the first $10\,\angstrom$ of available data. Within that $10\,\angstrom$ range, we computed the COS continuum as a linear combination of the LineTools continuum and the eBOSS mean continuum, beginning with 100\% and 0\% contributions from the COS and eBOSS continua on the red end, respectively, and ending with 0\% and 100\% contributions from the COS and eBOSS continua on the blue end of the available data. The remaining COS continuum in the region with missing data was then taken to be equal to the eBOSS mean continuum in that region. 

After completing this extrapolation, we had 48 COS continua with full coverage in the rest wavelength range of $1040-1600\,\angstrom$. However, low-$z$ quasars such as those in the COS dataset tend to be lower luminosity than typical DESI quasars. This is potentially a problem as there is an anti-correlation between the luminosity of the quasar continuum and the equivalent widths of its emission lines \citep{baldwin_luminosity_1977}. Because the COS quasars tend to be lower luminosity, they will feature stronger broad emission lines and therefore may not be representative of the DESI sample. In order to reduce the low-luminosity bias of our synthetic spectra, we removed the COS objects that are least similar to DESI Y1 quasars from our sample. 

\begin{figure}
    \centering
    \includegraphics[width=\columnwidth]{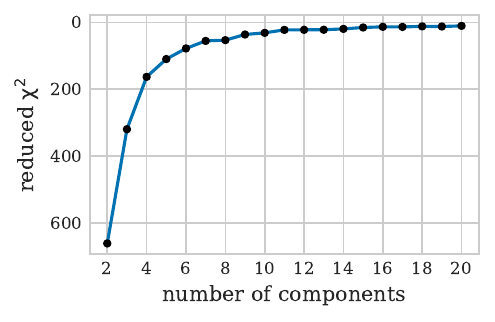}
    \caption{Reduced $\chi^2$ for NMF models with a number of components ranging from $2-20$. These components were derived from our initial sample of 48 COS continua. We used models with six components throughout this paper.}
    \label{fig:NMF_chi2}
\end{figure}

We identified the COS quasars that are least similar to the DESI quasars through an analysis of their spectra on the red side of the Ly$\alpha$ emission lines. First, we derived Nonnegative Matrix Factorization (NMF) components from our 48 COS continua. NMF is a dimensionality reduction technique similar to PCA, but the components are restricted to be nonnegative, which is more physically representative of quasar spectra. The components are not necessarily orthogonal as a consequence of this non-negativity constraint. We used the NMF code developed by \cite{zhu_nonnegative_2016} which includes the ability to use weights and mask pixels. We masked pixels where there was a $\geq 5\sigma$ difference between the LineTools continuum and flux, and derived a set of six NMF components from our COS continua. Figure~\ref{fig:NMF_chi2} shows the reduced $\chi^2$ curve for NMF models for $2-20$ components, which begins to flatten out at six components. We fit a linear combination of these six NMF components to the $1230-1600\,\angstrom$ region for both the 48 COS quasars and 10,000 relatively high-SNR (median $r-$band SNR $\geq 5$) DESI Y1 quasars. The six-dimensional parameter space of coefficients resulting from this fit represents the shape of DESI quasar spectra on the red side of Ly$\alpha$. 

\begin{figure*}
    \centering
    \includegraphics[width=\textwidth]{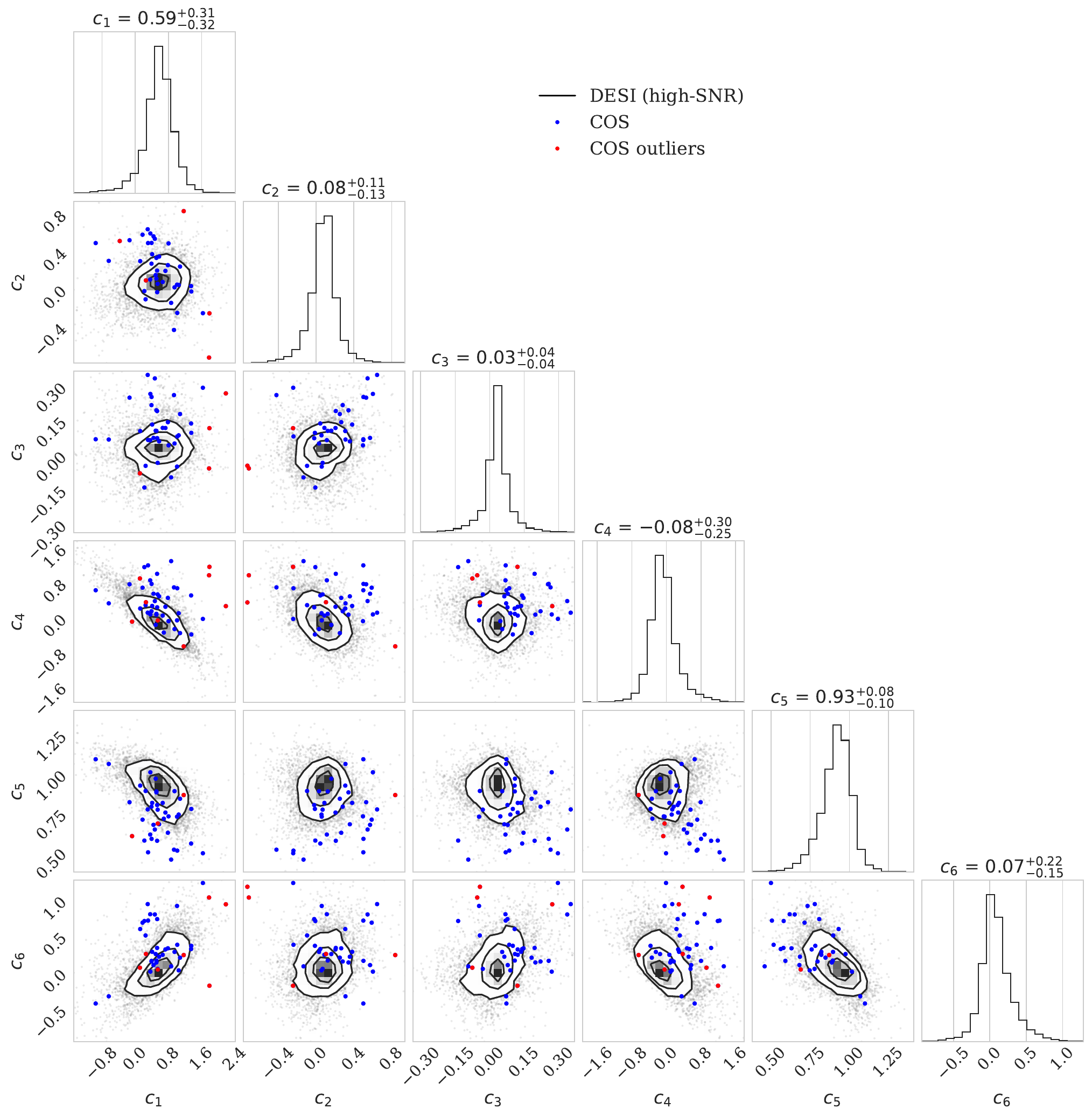}
    \caption{Distribution of coefficients obtained from fitting six NMF components to the $1230-1600\,\angstrom$ region of 10,000 DESI Y1 ({\it gray points and black contours}) and 48 COS quasar spectra ({\it large blue and red points}). The six components were derived from COS observations, where $c_i$ indicates the coefficient associated with component $i$. The red points mark the 10 COS objects with the lowest likelihood ($<10^{-10}$) of being described by the GMM representation of the DESI distribution. These 10 quasar spectra ({\it red points}) were removed as unrepresentative of DESI quasars. The remaining 38 ({\it blue points}) were retained for the analysis.}
    \label{fig:nmfcoeffs}
\end{figure*}

We fit a Gaussian Mixture Model (GMM) to the DESI coefficients and chose five classes based on the Bayesian Information Criterion (BIC). The distribution of coefficients is shown in Figure~\ref{fig:nmfcoeffs}. Ten of the COS quasars fall significantly outside both the observed coefficient distribution of points from DESI and the GMM. The ten that we eliminated all have probabilities $<10^{-10}$ of being described by the GMM. The COS quasars that we retained are marked with blue points in Figure~\ref{fig:nmfcoeffs} and the ones that we eliminated are marked with red points. 

\subsection{Synthetic Spectra}\label{sec:synthspec}
\begin{figure}
    \centering
    \includegraphics[width=\columnwidth]{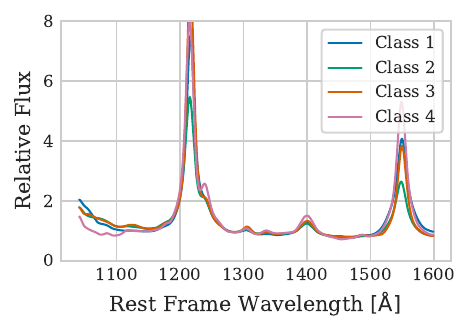}
    \caption{The mean spectra for our four classes of COS continua from the GMM representation of the NMF coefficients.}
    \label{fig:gmmclasses}
\end{figure}

We generated synthetic spectra based on a GMM of our final sample of COS continua. First, we removed two BL-Lac objects from our COS sample, then used the pixel masking criteria described in Section~\ref{sec:preprocess} to derive six new NMF components from the remaining sample of 36 masked COS continua. To create the synthetic spectra, we generated a GMM with four classes from the six-dimensional parameter space of NMF coefficients. We used the BIC to choose a four-component mixture model. The mean spectrum for each of these four classes is shown in Figure~\ref{fig:gmmclasses} and the main difference between the classes is the strength of the emission lines. For each of the four classes, we randomly selected an equal number of COS continua that are members of the corresponding class. Afterwards, we added a representative number of BL-Lac spectra based on the fraction of those present in the COS sample (two out of 38). The inclusion of BL-Lac-like objects allows our network to generalize better to DESI-like data with weaker emission line properties. Our sample of synthetic COS continua totals 20,000. 

This sample of continua forms the basis of our synthetic spectra. To create more diversity, we linearly perturbed each of these spectra according to 
\begin{equation}
    C_{\rm perturb}(\lambda) = C(\lambda) + M_0 + M_1 \left(\frac{\lambda/\angstrom - 1040}{1600-1040}\right)
\end{equation}
as in \cite{sun_quasar_2023} where $M_1$ was drawn from a uniform distribution from  $-0.2$ to $0.2$ and affects the slope of the continuum, and $M_0$ was drawn from a different uniform distribution and affects the strength of emission lines. To create COS-like synthetic spectra, we drew $M_0$ from a uniform distribution from $-0.1$ to $0.1$ so as to not systematically affect the emission line strengths. For another sample of 20,000 synthetic spectra that are more DESI-like, we drew $M_0$ from a uniform distribution from $2$ to $6$ based on visual inspection of DESI Y1 quasars. The higher values of $M_0$ reduce the strengths of the emission lines relative to the continuum, which is more representative of higher redshift, higher luminosity quasars. These linear perturbations add diversity to our synthetic spectra and increase the generalizability of our neural network.

After adding perturbations, we re-normalized each spectrum at 1280\,\angstrom. We then added $2-4$ random absorption features to the red side ($1216-1600\,\angstrom$) of our perturbed continua to mimic metal absorption lines. Each absorption line is a Voigt profile with a Lorentzian full width at half maximum (FWHM) of 0.1 and a Gaussian FWHM of either 0.4 for the COS-like spectra or 1.0 for the DESI-like spectra. The Lorentzian amplitude was randomly selected between $2-5$ for the COS-like spectra or $1.5-5$ for the DESI-like spectra. The range of absorption features and the parameter values were chosen based on visual inspection of the COS and DESI Y1 spectra.

Lastly, we added noise to each of these synthetic spectra. In order to incorporate a diverse SNR range into our synthetic sample, we added noise according to the SNR range observed in COS and DESI Y1 data. For the COS-like sample, we added Gaussian noise to each of the 20,000 synthetic spectra by randomly selecting error spectra belonging to the associated GMM class. We repeated this for the 20,000 DESI-like synthetic spectra using a sample of 40,000 DESI Y1 quasars that span the noise properties of the full Y1 sample.

Our final sample contains 40,000 COS-based synthetic spectra, of which 20,000 are COS-like in their emission line and noise properties, and the other 20,000 are DESI-like. This diversity in synthetic spectra helps ensure that our neural network learns to be robust against varying levels of noise, emission line strengths, and metal absorption line properties. The dispersion of our DESI-like COS-based synthetic spectra is shown in the second panel of Figure~\ref{fig:specdiversity}.

\section{Lyman-alpha Continuum Analysis Network}\label{sec:lycan}

In this section we present our neural network for predicting the quasar continuum, the Ly$\alpha$ Continuum Analysis Network (LyCAN). We first describe the dataset we used to train LyCAN in Section~\ref{sec:trainset}. This dataset is based on the COS-based synthetic spectra and DESI mocks described in the previous sections. We then describe how we optimized the hyperparameters for LyCAN and the network architecture in Section~\ref{sec:network}. Lastly we quantify the performance of our network and compare it to that of PCA- and NMF-based prediction models in Section~\ref{sec:performance}.

\subsection{Training Set}\label{sec:trainset}
The goal of our work is to predict continua for DESI quasars. Additionally, we want to avoid biasing our network towards low-luminosity quasars, which are over-represented in the COS dataset. We combined our 40,000 COS-based synthetic spectra with a sample of 40,000 DESI Y5 mock spectra with true continua. These 80,000 spectra are the input data for our neural network. We performed a randomized 65\%/25\%/10\% split to select our training, testing, and validation sets, respectively. The testing set was never seen during training and is only used to quantify the performance of our model afterwards, while the validation set was used during the training process to evaluate the mean squared error loss function at each epoch, allowing for the determination of optimal weights.

\begin{figure*}
    \centering
    \includegraphics{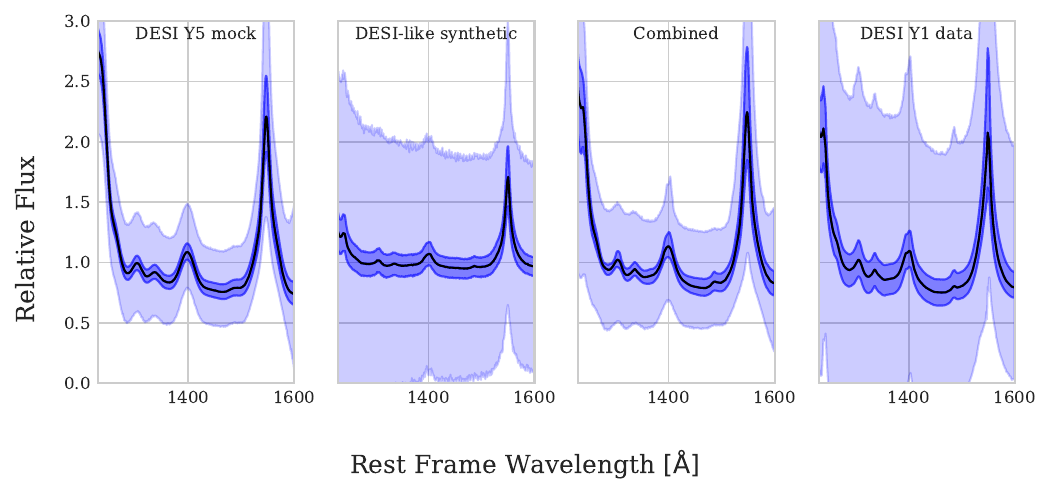}
    \caption{Illustration of the diversity of the spectra. From left to right: DESI Y5 mock spectra included in our input data, COS-based synthetic spectra with DESI Y1-like noise, the combination of Y5 mock spectra and COS-based synthetic spectra with COS-like and DESI Y1-like noise, and the real DESI Y1 data. The shaded areas indicate the $1-$ and $2-\sigma$ regions. The third panel is representative of our training set.}
    \label{fig:specdiversity}
\end{figure*}

The diversity of these spectra is shown in Figure~\ref{fig:specdiversity}. The figure shows these samples before the training/testing/validation split, but each class of spectra is equally represented in each of these subsets. The third panel shows the combination of DESI Y5 mock spectra and COS-based synthetic spectra with COS-like and DESI Y1-like noise, which is representative of our training set.

\subsection{Network}\label{sec:network}
We created the Convolutional Neural Network (CNN) model LyCAN to predict the forest continuum. A neural network model is a good choice for this application  due to its ability to learn nonlinear representations of the data, unlike PCA- and NMF- based methods. A CNN is a type of feedforward neural network with one or more convolutional layers that convolve the input pixels with learned filters. These layers are designed to capture hierarchical features in the input data. Optimizable hyperparameters in these convolutional layers include the number of filters and kernel size, among others. Dense (fully connected) layers are often added to the latter part of the network to combine high-level features learned by the convolutional layers. The number of nodes per dense layer is another optimizable hyperparameter.

We optimized the CNN architecture with a pseudo-random search in which we varied the number of convolutional layers, the kernel size and number of filters for these layers, the number of dense layers, and the number of nodes per dense layer. The number of convolutional layers ranged from one to four (including the input layer), and dense layers also ranged from one to four (not including the output layer). We trained and tested 10 different models for each possible combination (e.g. one convolutional layer and one hidden dense layer for the least complex models, up to four convolutional layers and four hidden dense layers for the most complex models). Each of these models was generated by randomly selecting the number of filters per convolutional layer and the number of nodes per hidden dense layer from a range of options. The number of filters was chosen to be a multiple of 16 between $16-64$ and the number of nodes was chosen to be a multiple of 32 between $32-256$. The number of nodes in the output dense layer is always equal to the number of wavelength pixels in the full spectral range\footnote{Due to edge effects from resampling, the rest frame wavelength range actually spans from 1040.8-1599.2~\angstrom~(inclusive).}. We separately trained these 160 configurations with a kernel size of three and a kernel size of six for a total of 320 unique neural network architectures. 

\begin{figure}
    \centering
    \includegraphics[width=\columnwidth]{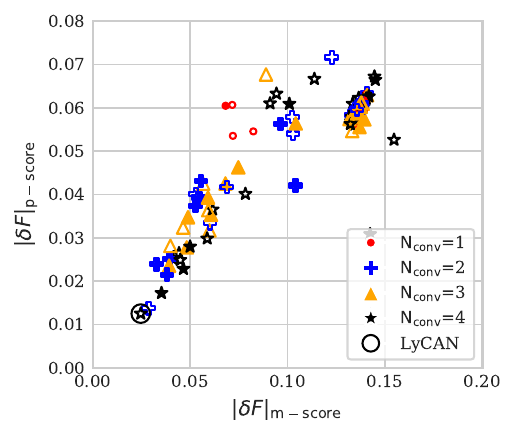}
    \caption{AFFE scores for candidate CNN architectures. Filled symbols indicate that a kernel size of three was used for the convolutional layers, while open symbols indicate a kernel size of six. The colors and symbols of the points vary depending on the number of convolutional layers in the network. Our chosen architecture is marked with a large circle in the lower left and minimizes both the median AFFE score ($\left\lvert\delta \rm F \right\rvert_\text{m-score}$; Eq.~\ref{affemscore}) and the $1\sigma$ range AFFE score ($\left\lvert\delta \rm F \right\rvert_\text{p-score}$; Eq.~\ref{affepscore}).}
    \label{fig:cnnopt}
\end{figure}

We trained each of these unique architectures with the training subset of our data, and quantified each model's performance on the testing subset. Among the testing subset, we evaluated the performance on the COS-based synthetic and DESI Y5 mock spectra with the Absolute Fractional Flux Error (AFFE), defined as 
\begin{equation}\label{affe}
\left\lvert\delta F \right\rvert = \frac{\int_{\lambda_1}^{\lambda_2} \left\lvert \frac{F_{\rm pred}(\lambda) - F_{\rm true}(\lambda)}{F_{\rm true}(\lambda)} \right\rvert d\lambda}{\int_{\lambda_1}^{\lambda_2} d\lambda},
\end{equation}
where $F_{\rm true}$ is the true continuum, $F_{\rm pred}$ is the predicted continuum, $\lambda_1=1040\,\angstrom$, and $\lambda_2=1200\,\angstrom$. We compared the resulting AFFE values in the forest region for each trained model by calculating a median AFFE score per model of
\begin{equation}\label{affemscore}
    \left\lvert\delta \rm F \right\rvert_\text{m-score} = \sqrt{\left\lvert\Tilde{\delta \rm F} \right\rvert_{\rm S}^2 + \left\lvert\Tilde{\delta \rm F} \right\rvert_{\rm M}^2}
\end{equation}
where $\lvert\Tilde{\delta \rm F} \rvert_{\rm S}$ refers to the median AFFE for the COS-based synthetic testing sample and $\lvert\Tilde{\delta \rm F} \rvert_{\rm M}$ refers to the median AFFE for the DESI mock testing sample. In addition, we evaluated another metric based on the percentile range of AFFEs,
\begin{equation}\label{affepscore}
    \left\lvert\delta \rm F \right\rvert_\text{p-score} = \sqrt{\left\lvert\delta \rm F \right\rvert_{\rm S,1\sigma}^2 + \left\lvert\delta \rm F \right\rvert_{\rm M,1\sigma}^2},
\end{equation}
where $\lvert\delta \rm F\rvert_{\rm S,1\sigma}$ denotes the $1\sigma$ AFFE range for the COS-based synthetic testing sample and $\lvert\delta \rm F \rvert_{\rm M,1\sigma}$ is the $1\sigma$ AFFE range for the DESI mock testing sample. Figure~\ref{fig:cnnopt} shows the metrics in Equations \eqref{affemscore} and \eqref{affepscore} for a selection of the tested models.

We chose the architecture that minimized the median AFFE and $1\sigma$ range scores. This architecture features six hidden layers, three of which are convolutional layers and the remaining three are dense layers. The full architecture is shown in Table~\ref{tab:arch}. LyCAN was compiled and trained using Keras\footnote{\url{https://keras.io/}} with the TensorFlow\footnote{\url{https://www.tensorflow.org/}} backend. We used the linear activation function for the output layer and the rectified linear unit \citep[ReLU;][]{relu} activation function for all other layers, the mean squared error loss function, and the Adam optimizer \citep{kingma2017adammethodstochasticoptimization}. We trained for a maximum of 50 epochs and implemented early stopping criteria that monitored the validation loss starting at epoch 20 with a patience of five and a minimum delta of $10^{-5}$. The patience sets the number of epochs with negligible improvement after which training is stopped and optimal weights are restored, while the delta sets the minimum change in the validation loss to qualify as an improvement. There are a total of 169,851 trainable parameters (i.e. weights and biases) in our network. Lastly, the output continua were smoothed using a Gaussian filter with a standard deviation of three in order to reduce the noise in the predictions and minimize the AFFEs.

We tested other versions of CNNs for this work. For example, we trained a CNN with the same architecture on only mock spectra and tested it on each of the same testing sets. We found no improvement in the performance on mock spectra, and a significant reduction in performance on our COS-based synthetic spectra, including the set with DESI-like emission line and noise properties. We also trained a CNN on only a sample of COS-based synthetic spectra with COS-like emission line and noise properties, and found equally good performance on the testing set for the same type of objects, but markedly poorer performance on the COS-based synthetic spectra with DESI-like properties and the DESI mock spectra. Finally, we trained a CNN on only COS-based synthetic spectra containing an equal representation of COS- and DESI-like properties. Median AFFEs on the COS-based synthetic samples remained low, and a moderate median AFFE of $\sim0.08$ on the DESI mock spectra was achieved. This shows that the addition of COS-based synthetic spectra with DESI-like emission line and noise properties helps the network more accurately learn how to predict continua for DESI-like objects. Ultimately, a combination of COS-based synthetic spectra with varying emission line and noise properties and DESI mock spectra proved to be most effective.

\begin{table}
	\centering
	\caption{LyCAN architecture. The first column specifies whether the layer is a convolutional layer, a max pooling layer (occurring after each convolutional layer), or a dense (fully connected) layer. The size column specifies the number of filters for convolutional layers or neurons for dense layers. In the case of the max pooling layers, pixels are pooled in groups of two. We use the linear activation function for the output layer, and the ReLU activation function for all other layers where applicable. The number of nodes in the output layer is equal to the number of pixels in the predicted continuum. The kernel size is six for the convolutional layers.}
	\label{tab:arch}
	\begin{tabular}{lcr}
		\hline
		Layer type & Size & Activation function\\
		\hline
		Conv1D (input) & 64 & ReLU\\
        MaxPooling1D & 2 & \\
		Conv1D & 32 & ReLU\\
        MaxPooling1D & 2 & \\
        Conv1D & 48 & ReLU\\
        MaxPooling1D & 2 & \\
        Conv1D & 16 & ReLU\\
        MaxPooling1D & 2 & \\
        Flatten &  & \\
		Dense & 96 & ReLU\\
        Dense & 64 & ReLU\\
        Dense & 128 & ReLU\\
        Dense (output) & 699 & Linear\\
		\hline
	\end{tabular}
\end{table}

\subsection{Performance and Comparisons}\label{sec:performance}

We quantify the performance of our network with the AFFE metric defined in Equation~\eqref{affe}. We also used its non-absolute counterpart, the Fractional Flux Error (FFE), to study systematic biases in the LyCAN continuum predictions.

\begin{figure*}
    \centering
    \includegraphics{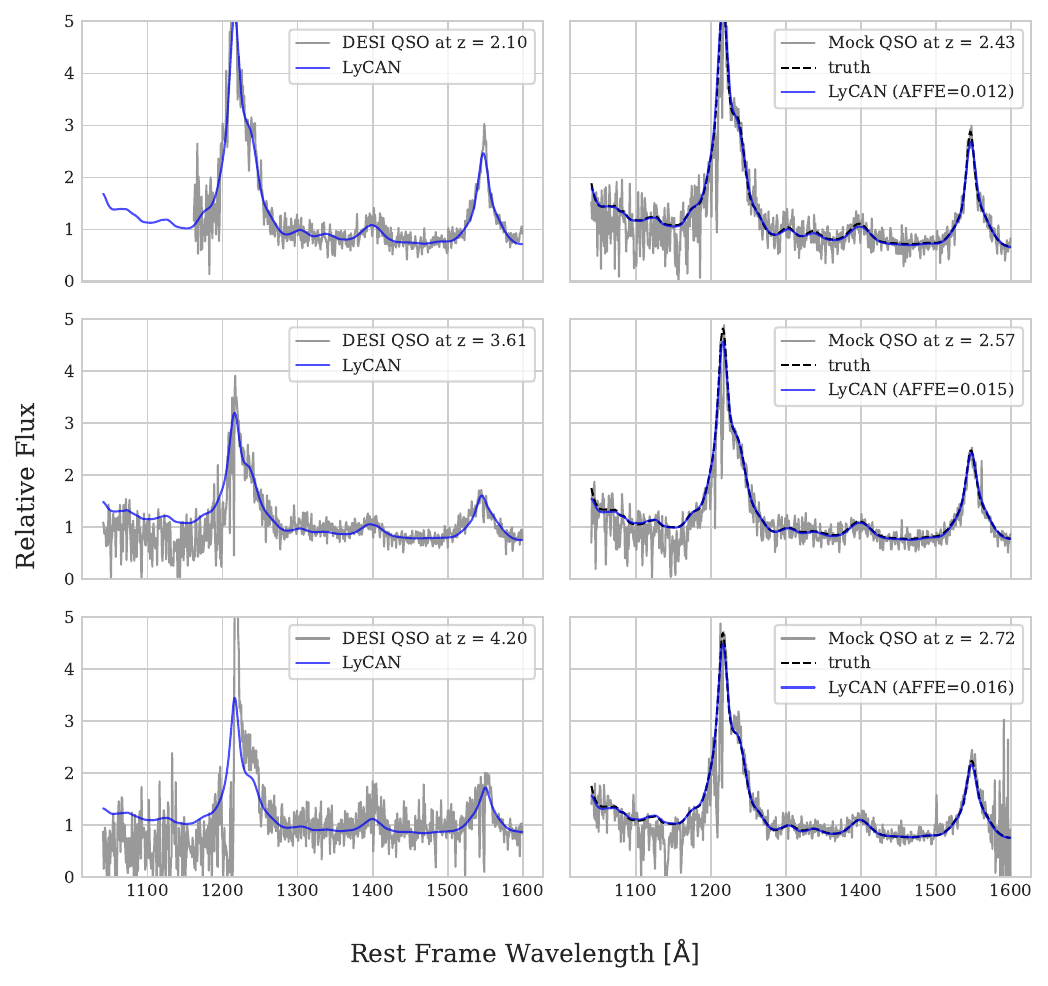}
    \caption{LyCAN continuum predictions for DESI Y1 and DESI Y5 mock quasars. The panels on the left show predictions for real DESI Y1 quasars at increasing redshifts from top to bottom, and the right panels show predictions on DESI Y5 mock spectra. The panels with mock spectra also include their true continua and the AFFE values in the forest region.}
    \label{fig:desipreds}
\end{figure*}

Figure~\ref{fig:desipreds} shows continuum predictions for real DESI Y1 QSOs and Y5 mock QSOs in our testing set at several redshifts. The truth continua for the mock spectra are also shown, but are difficult to distinguish because the LyCAN predictions are nearly identical. AFFE values are shown in the legend for the mock spectra.

We created PCA- and NMF-based prediction models to compare the performance of these two other well-known methods to LyCAN. These models are based on the same training data as LyCAN, and are distinct from the previous NMF models developed to remove COS outliers and help generate COS-based synthetic spectra. We used scikit-learn\footnote{\url{https://scikit-learn.org/stable/modules/generated/sklearn.decomposition.PCA.html}} to derive PCA components from the training set continua after z-score standardization. This scaled each spectrum so that its mean continuum is zero and standard deviation is one, which reduces possible biases due to luminosity. Once we computed the PCA continuum predictions, we reversed the standardization. 
We optimized the PCA and NMF models by testing models with up to $16$ components. For each model, we fit the components to the red side of the training data ($1216-1600\,\angstrom$) and then used the best-fit coefficients to extrapolate the continuum from the red side to the full spectral range, including the forest. We used the median AFFE on the DESI mock testing set as a performance metric to find the optimal models. For both PCA and NMF, we found that four components minimized this metric, and we therefore adopted these as the optimal models.

\begin{figure*}
    \centering
    \includegraphics[width=\textwidth]{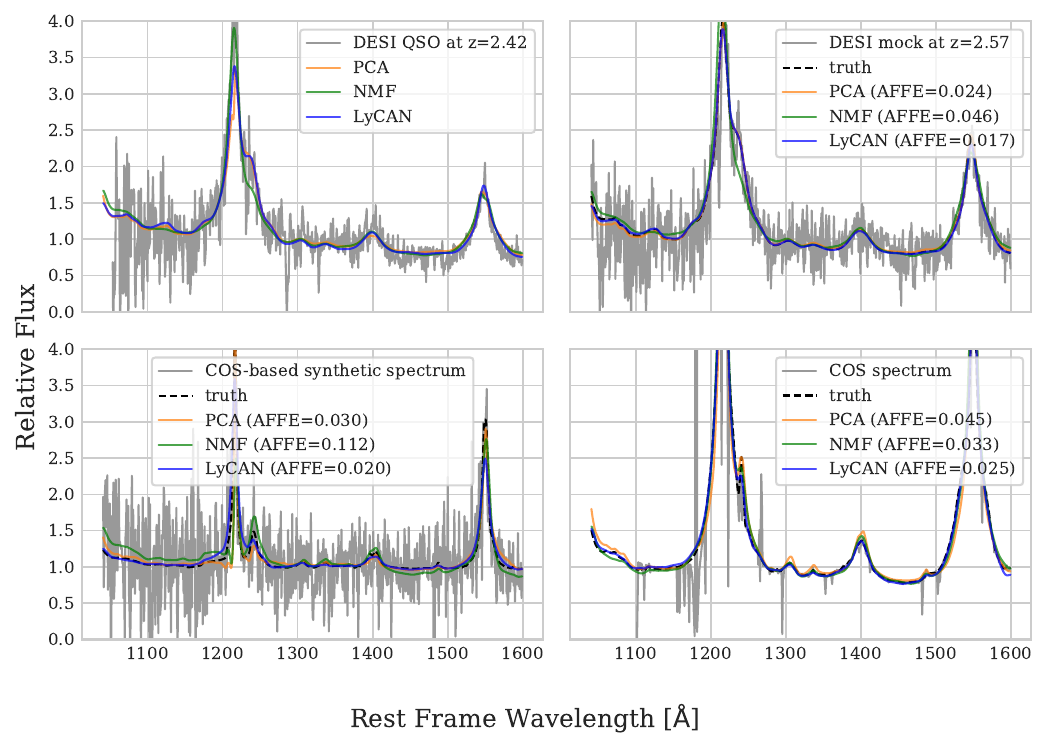}
    \caption{Examples of PCA, NMF, and LyCAN continuum predictions for four different classes of spectra: a real DESI Y1 spectrum, a DESI Y5 mock spectrum, a COS-based synthetic spectrum, and a real COS spectrum. None of these spectra appear in our training set.}
    \label{fig:predgrid}
\end{figure*}

Figure~\ref{fig:predgrid} shows a comparison of the PCA and NMF continuum predictions to LyCAN for four types of spectra. Neither technique works as well as LyCAN. In addition to their poorer performance, the components of the PCA and NMF models must be fit to smoothed continua in order to yield reasonable results. We therefore used the red side of the continua output by LyCAN as input to our PCA- and NMF-based models when predicting continua for DESI Y1 quasars.

\begin{figure*}
    \centering
    \includegraphics{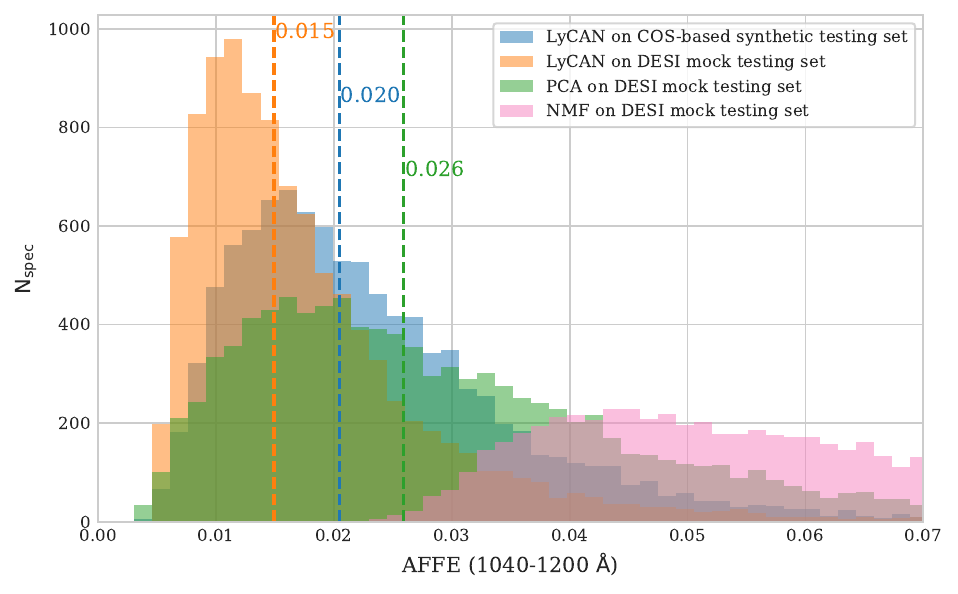}
    \caption{Distribution of Absolute Fractional Flux Errors (AFFEs) in the forest region for LyCAN, PCA, and NMF models on the DESI mock testing set. We also show the distribution of AFFEs for LyCAN predictions on our COS-based synthetic testing spectra. The dashed vertical lines indicate the median AFFE for each sample.}
    \label{fig:affehist}
\end{figure*}

LyCAN outperforms these PCA- and NMF- based models. Figure~\ref{fig:affehist} shows the distribution of AFFE values for different models and testing samples. The median AFFE values for different datasets and models are shown in Table~\ref{tab:affes}. The first column shows the median AFFE values for these three models on the real COS quasars used to generate the synthetic data in the next two columns. LyCAN is able to predict the forest continua for the COS sample more accurately than the NMF and PCA models, although this sample is quite small. The next two columns of Table~\ref{tab:affes} list the median AFFEs of each model on the synthetic training and testing spectra, respectively, generated from the COS objects. Both of these subsets include synthetic spectra with DESI- and COS-like noise. LyCAN again outperforms the PCA and NMF models, achieving a median AFFE of 2.0\%. We see nearly identical performance on the training versus testing subsets across all three models. Finally, the last two columns of Table~\ref{tab:affes} show the median AFFEs from each prediction model on the DESI Y5 mock training and testing sets. Again, LyCAN performs the best out of all the models, achieving a median AFFE of 1.5\%, approximately 40\% better than the performance of the PCA model on the same dataset (2.6\%) and almost five times better than the performance of the NMF model (7.4\%). Interestingly, the NMF model performs more poorly than the PCA model in all cases except for the real COS sample. This is a surprising result, and may indicate that the NMF components are not as generalizable to more diverse datasets as the PCA components, whether that is due to the number of components used, the components themselves, or the standardization process. We also note that while the number of components has been optimized for our PCA and NMF models, more sophisticated PCA-based models in the literature that use pixel weighting and/or a projection matrix approach may offer better performance \citep[e.g.][]{suzuki_predicting_2005,paris_principal_2011,lee_mean-flux_2012,davies_predicting_2018}.

\begin{table*}
	\centering
    \setlength\tabcolsep{1pt}
	\caption{A comparison of the median Absolute Fractional Flux Errors for PCA, NMF, and LyCAN prediction models on different samples.}
	\label{tab:affes}
	\resizebox{\textwidth}{!}{\begin{tabular}{lccccr}
		\toprule
        Model & COS sample & \thead{COS-based synthetic\\ sample, training} & \thead{COS-based synthetic\\ sample, testing} & \thead{DESI mock sample, \\ training} & \thead{DESI mock sample, \\ testing}\\
        \midrule
		LyCAN & 0.041 & 0.020 & 0.020 & 0.015 & 0.015 \\
        NMF & 0.066 & 0.138 & 0.137 & 0.074 & 0.074 \\
		PCA & 0.069 & 0.039 & 0.039 & 0.026 & 0.026 \\
		\bottomrule
	\end{tabular}}
\end{table*}

\begin{figure*}
    \centering
    \includegraphics{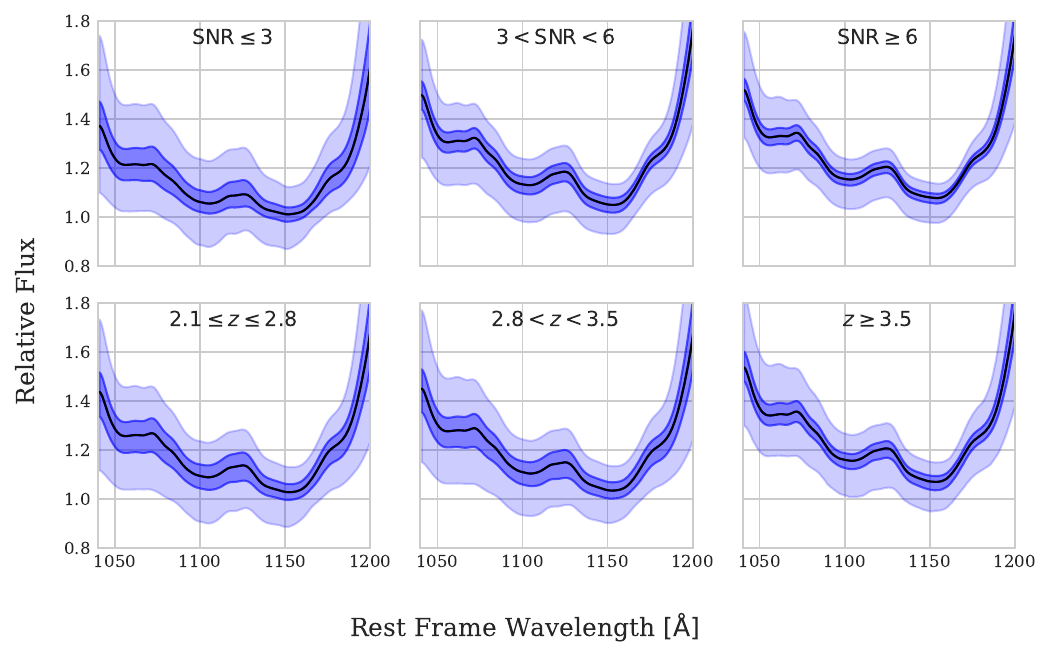}
    \caption{The dispersions of continuum predictions for DESI Y1 quasars. The top panel shows three ranges of SNR, while the bottom panel shows three ranges of redshift. The solid line in each panel corresponds to the median prediction for that SNR or redshift range, and the shaded regions indicate the $1-$ and $2-\sigma$ regions.}
    \label{fig:desi_pred_disp}
\end{figure*}

We investigated the dependence of our DESI Y1 LyCAN continuum predictions on redshift and SNR. Figure~\ref{fig:desi_pred_disp} shows the dispersion of continuum predictions for DESI Y1 quasars as a function of SNR and redshift. Overall, there is slightly less dispersion in the high-SNR predictions, while the median continuum remains consistent. We see a smaller but similar effect in the dispersions of the higher-redshift predictions. The shape of the predictions remains consistent in each SNR and redshift range.

\begin{figure*}
    \centering
    \includegraphics{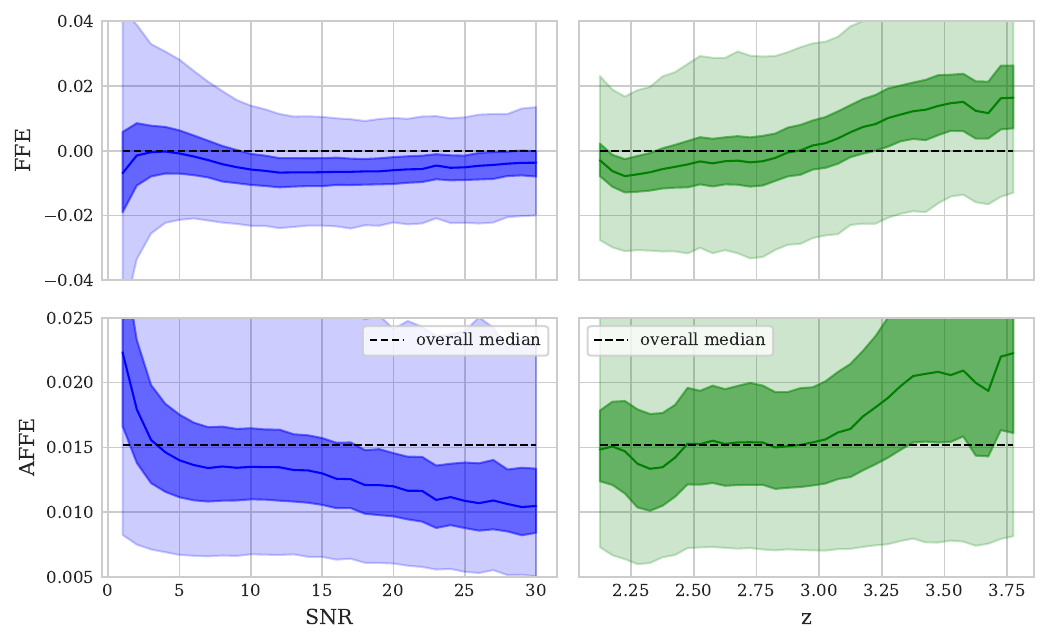}
    \caption{Systematic biases of LyCAN predictions on DESI Y5 mock spectra. The top panels show the Fractional Flux Error as functions of SNR and $z$, while the bottom panels show the Absolute Fractional Flux Error (see Eq. \ref{affe}) as functions of SNR and $z$. The solid line indicates the median (A)FFE, and the shaded regions show the $1-$ and $2-\sigma$ regions. LyCAN tends to perform best on higher SNR, low-$z$ spectra, and is more likely to over-predict the true continuum at higher $z$. This is a small effect ($\sim2$\% at highest $z$).}
    \label{fig:mock_systematics}
\end{figure*}

Similarly, we investigated the dependence of both FFE and AFFE for DESI Y5 mock continuum predictions on the redshift and SNR of the quasar. This provides an estimate of any bias in LyCAN that may produce nonphysical trends with redshift and/or SNR. Figure~\ref{fig:mock_systematics} shows these systematic biases. LyCAN performs more accurately on low to intermediate redshift, higher SNR objects. Based on the FFE, LyCAN slightly over-predicts the continuum level at higher redshifts by $\sim1$\%. We correct for this effect in Section~\ref{sec:tau}.

\section{Evolution of the effective optical depth}\label{sec:tau}

In this section we use the DESI Y1 quasar continuum predictions from LyCAN to measure the evolution of the effective optical depth of the Lyman-$\alpha$ forest. We detail our optical depth calculation in Section~\ref{sec:taucalc}. This calculation is a measurement of the total optical depth in the forest region, which includes contributions from metals and optically thick absorbers, in addition to the Lyman-$\alpha$ forest. In Section~\ref{sec:taucorrs} we describe how we correct for these contributions, as well as quantify the potential bias in our continuum prediction, to determine the optical depth due to the Lyman-$\alpha$ forest. Lastly, we compare our DESI Y1 results to previous results from the literature in Section~\ref{sec:taucomparisons}.

\subsection{Calculation}\label{sec:taucalc}

LyCAN enables an absolute measurement of the effective optical depth, as it does not use any spectral information in the forest region, and we make this measurement with the largest sample to date. We do not use every DESI quasar for our measurement since we require a sufficiently high SNR to accurately measure the absorption present in the forest region. We use DESI Y1 quasars with median $r-$band SNR $\geq$ 5 and remove objects with BAL features, but retain DLAs because we later correct our measurements for optically thick absorbers. Our choice of SNR cut and BAL removal is consistent with previous works \citep[e.g.][]{kamble_measurements_2020}. Among the full sample of $\sim450,000$ $z\geq2.1$ DESI Y1 quasars, roughly 90,000 contain BALs and $\sim14,000$ have redshift warnings. The largest cut is due to the minimum SNR requirement. Our final sample contains 83,635 DESI Y1 quasars.

For each quasar, we measure the optical depth in redshift bins of width $\Delta z_{\rm Ly\alpha} = 0.1$ between $2.0 \leq z_{\rm Ly\alpha} \leq 4.2$, where $z_{\rm Ly\alpha} \equiv \lambda_{\rm obs}/\lambda_{\rm Ly\alpha}-1$ is the redshift of the Ly$\alpha$ pixel in the forest region. We restrict the analysis to the rest frame wavelength range of $1070-1160\,\angstrom$ in each quasar as in \cite{kamble_measurements_2020} to mitigate contamination from the Ly$\alpha$ and \ion{O}{6} emission lines. In each redshift bin, we compute the 3$\sigma$-clipped mean transmission as
\begin{equation}
    \langle F \rangle = \left\langle \frac{f(z_{\rm Ly\alpha})}{C_q(z_{\rm Ly\alpha})} \right\rangle
\end{equation}
where $f$ is the observed flux and $C_q$ is the LyCAN prediction of the unabsorbed continuum. This $\rm N_{\rm QSO}\times \rm N_{z}$ matrix is our measurement of the mean transmission per quasar in each redshift bin. We convert this to an optical depth matrix according to $\tau=-\ln\langle F\rangle$ and compute the 3$\sigma$-clipped mean $\tau_{\rm eff}$ per redshift bin. These raw measurements are listed in Table~\ref{tab:tau}.

\begin{figure*}
    \centering
    \includegraphics[width=0.95\textwidth]{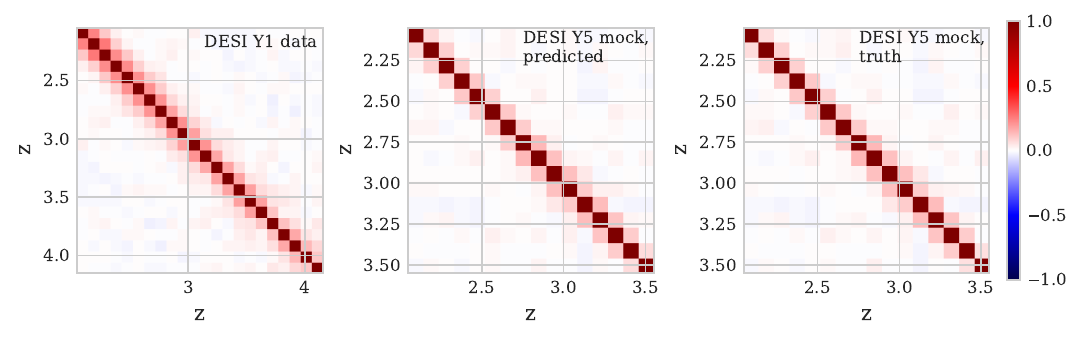}
    \caption{Correlation matrices for optical depth measurements in DESI Y1 with LyCAN continuum predictions, Y5 mock spectra with LyCAN predictions, and the same sample of Y5 mock spectra with their true continua. These matrices do not include corrections for uncertainties due to continuum bias, metals, or optically thick absorbers.}
    \label{fig:corrs}
\end{figure*}

We quantify our statistical uncertainties with 10,000 bootstrap realizations to compute the covariance matrix. In this calculation we randomize the quasars in the sample and compute the 3$\sigma$-clipped mean optical depth per redshift bin across all objects in the sample. There is covariance in our measurements because most quasars contribute to the optical depth measurement at several redshifts. The correlation matrix is shown in Figure~\ref{fig:corrs} and the diagonal of the covariance matrix $\sigma_\text{stat}$ is listed in Table~\ref{tab:tau}.

We repeat this analysis with the DESI Y5 mock spectra, once with the LyCAN continuum predictions and once with the true continua from the mocks. The correlation matrices for each of these datasets are also shown in Figure~\ref{fig:corrs}. The mock quasars do not span the full redshift range of the data, although the binning in redshift is the same. While the measurements on DESI Y1 data show correlations over several bins that is consistent with the redshift range spanned by individual forests, there are only some modest correlations in the immediately adjacent redshift bin for both of the analyses on mocks. This is likely because LyCAN performs better on DESI mocks than DESI data.  

\subsection{Corrections and Systematic Error}\label{sec:taucorrs}

\begin{figure}
    \centering
    \includegraphics[width=\columnwidth]{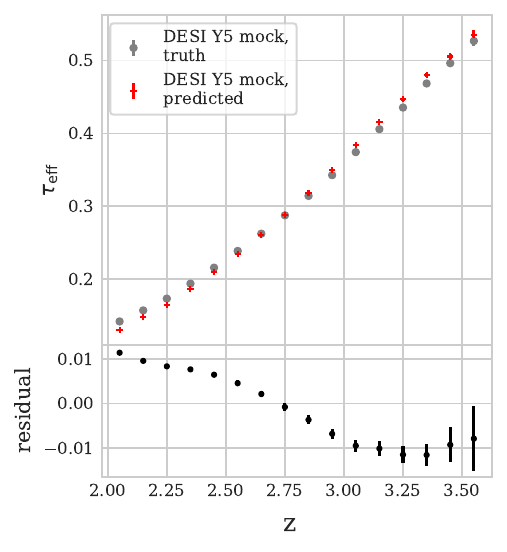}
    \caption{Top: effective optical depth measurements for the DESI Y5 mock dataset based on truth continua (\textit{gray points}) or LyCAN continuum predictions (\textit{red crosses}). These are raw measurements, not corrected for metals or optically thick absorbers. The error bars represent the statistical uncertainties. Bottom: the residual between the true and predicted measurements with statistical error bars.}
    \label{fig:tau_mock}
\end{figure}

Our measurement of the total optical depth described in \S\ref{sec:taucalc} includes other absorption features in addition to Ly$\alpha$, most notably metal lines and optically thick absorbers. There are also small biases in the continuum predictions, which are illustrated in Figure~\ref{fig:mock_systematics}. We correct the total optical depth measurement for both the continuum biases and other absorption features to determine the evolution of the optical depth of the Ly$\alpha$ forest. We first use the DESI Y5 mock dataset to calibrate our corrections for the redshift bias in the LyCAN continuum predictions. Figure~\ref{fig:tau_mock} shows our effective optical depth measurements for the mock dataset based on the LyCAN predicted continua and the mock truth continua. We attribute the difference to the continuum bias, which may be due to the systematic selection of higher luminosity quasars at higher redshifts. The top right panel in Figure~\ref{fig:mock_systematics} shows that LyCAN tends to slightly under-predict the true continuum at lower redshifts and slightly over-predict the continuum at higher redshifts. This effect is small ($\sim1$\% in FFE) and is consistent with what is shown in Figure~\ref{fig:tau_mock} -- the under-predictions at lower redshifts lead to less measured absorption, while the over-predictions at high redshifts lead to greater measured absorption than truth. Since the mock quasar samples in Figure~\ref{fig:tau_mock} are identical, the difference in continuum solely determines the observed discrepancy in measured optical depth. 

Because the FFE is a fractional error, it has an additive effect on the effective optical depth,
\begin{equation}\label{biascorrect}
    \tau_\text{eff, measured} = \tau_\text{eff, true} + \delta F,
\end{equation}
where $\tau_\text{eff, measured}$ is the measured optical depth with LyCAN continuum predictions, $\tau_\text{eff, true}$ is the true optical depth, and $\delta F$ is the fractional flux error (i.e. the non-absolute version of the AFFE; see Equation~\ref{affe}). We therefore use Equation~\ref{biascorrect} with the 3$\sigma$-clipped mean FFE per $z_{\rm Ly\alpha}$ bin to correct for this redshift bias and extrapolate to $z_{\rm Ly\alpha}>3.6$ from the highest redshift measurements in our mock sample. We also compute the uncertainty on the mean FFE, $\sigma_\text{bias}$, to quantify the uncertainty in this correction. Our bias-corrected $\tau_{\rm eff}$ measurements are reported in Table~\ref{tab:tau}. The uncertainties on this correction are negligible ($\sim10^{-4}$) compared to other sources of uncertainty and therefore not listed.

After correcting our DESI Y1 optical depth measurements for the LyCAN continuum bias, we apply corrections due to absorption from metals and optically thick absorbers. \citet{faucher-giguere_direct_2008} presented additive corrections due to metal absorption lines per redshift bin using the \cite{schaye_metallicity_2003} correction scheme. They also calculated a systematic error for this metal correction by examining the deviation between the \citet{schaye_metallicity_2003} and \cite{kirkman_h_2005} metal correction schemes. We adopt the metallicity correction from \citet{schaye_metallicity_2003} and the associated systematic errors $\sigma_{\rm \tau_{eff}, \text{metal}}$ from \citet{faucher-giguere_direct_2008}. The systematic error due to the metal correction scheme dominates our total error budget.

\citet{becker_refined_2013} presented a functional form for the flux decrement due to optically thick absorbers ($N_{\rm HI}\geq10^{17.2}~\text{cm}^{-2}$) which is dominated by super Lyman limit systems (SLLSs; $10^{19.0}\leq N_{\rm HI} < 10^{20.3}~\text{cm}^{-2}$) and DLAs ($N_{\rm HI}\geq10^{20.3}~\text{cm}^{-2}$). They found that the flux decrement is equal to $F_T=1-0.0045[(1+z)/3]^{3.0}$. We apply this to our effective optical depth measurements per redshift bin by subtracting $\tau_T=-\text{ln}(F_T)$ from our metal-corrected measurements. The authors did not report an associated uncertainty for this correction, so this is not included in our total error budget.

Our total uncertainties include statistical and systematic contributions which we add in quadrature. These are $\sigma_{\rm \tau_{eff, tot}} = \sqrt{\sigma_{\rm \tau_{eff, stat}}^2 + \sigma_{\rm \tau_{eff, bias}}^2 + \sigma_{\rm \tau_{eff, metal}}^2}$. As stated earlier, $\sigma_{\rm \tau_{eff, bias}}$ has a negligible effect on the total error budget, while $\sigma_{\rm \tau_{eff, metal}}$ dominates it. The $\sigma_{\rm \tau_{eff, tot}}$ values per redshift bin are given in Table~\ref{tab:tau}.

We fit our measurements with a power-law of the form $\tau(z)=\tau_0(1+z)^\gamma$. While we used the raw covariance matrix to quantify our statistical uncertainties (see \S\ref{sec:taucalc}), we computed a corrected covariance matrix to use as an input to the fitting procedure. This corrected covariance matrix was generated via bootstrapping as in Section~\ref{sec:taucalc} from a corrected per-quasar optical depth matrix. The corrections we used are not unique to quasars and only change with redshift bin. We find $\tau_0 = (2.46\pm0.14)\times 10^{-3}$ and $\gamma=3.62\pm0.04$. We report our bias-corrected and fully corrected measurements alongside the raw measurements in Table~\ref{tab:tau}. Our fully corrected measurements with total uncertainty error bars $\sigma_{\rm \tau_{eff, tot}}$ and the associated power-law fit are shown in Figure~\ref{fig:taueff}.

\begin{table*}
    \centering
	\caption{Effective optical depth measurements $\tau_{\rm eff}$ and associated uncertainties. $z_c$ refers to the redshift at the center of the bin.}
	\label{tab:tau}
    \begin{threeparttable}
	\begin{tabular}{lccccr}
        \toprule \midrule
		$z_c$ & Raw\tnote{a} & Bias-corrected\tnote{b} & Final\tnote{c} & $\sigma_{\tau_\text{\rm eff, stat}}$\tnote{d} & $\sigma_{\tau_\text{\rm eff, tot}}$\tnote{e} \\
        \midrule
        2.05 & 0.159 & 0.171 & 0.147 & 0.001 & 0.012\\
		2.15 & 0.174 & 0.183 & 0.158 & 0.001 & 0.012\\
        2.25 & 0.201 & 0.210 & 0.179 & 0.001 & 0.015\\
		2.35 & 0.224 & 0.232 & 0.200 & 0.001 & 0.016\\
        2.45 & 0.251 & 0.258 & 0.226 & 0.001 & 0.016\\
        2.55 & 0.267 & 0.272 & 0.235 & 0.001 & 0.018\\
        2.65 & 0.305 & 0.307 & 0.268 & 0.002 & 0.019\\
        2.75 & 0.333 & 0.333 & 0.292 & 0.002 & 0.020\\
        2.85 & 0.363 & 0.360 & 0.316 & 0.003 & 0.021\\
        2.95 & 0.394 & 0.388 & 0.342 & 0.003 & 0.022\\
        3.05 & 0.431 & 0.423 & 0.373 & 0.003 & 0.023\\
        3.15 & 0.468 & 0.459 & 0.410 & 0.003 & 0.023\\
        3.25 & 0.515 & 0.505 & 0.455 & 0.004 & 0.022\\
        3.35 & 0.562 & 0.552 & 0.498 & 0.005 & 0.025\\
        3.45 & 0.599 & 0.590 & 0.527 & 0.006 & 0.030\\
        3.55 & 0.654 & 0.646 & 0.579 & 0.007 & 0.032\\
        3.65 & 0.713 & 0.706 & 0.638 & 0.008 & 0.031\\
        3.75 & 0.771 & 0.764 & 0.694 & 0.009 & 0.032\\
        3.85 & 0.848 & 0.841 & 0.770 & 0.010 & 0.033\\
        3.95 & 0.908 & 0.901 & 0.830 & 0.016 & 0.034\\
        4.05 & 0.935 & 0.927 & 0.854 & 0.019 & 0.036\\
        4.15 & 1.011 & 1.003 & 0.928 & 0.023 & 0.039\\
        \bottomrule \midrule
	\end{tabular}
    \begin{tablenotes}\footnotesize
    \item[a] Raw $\tau_{\rm eff}$ measured directly from LyCAN continuum predictions
    \item[b] Raw measurement corrected for redshift-dependent continuum bias in LyCAN
    \item[c] Final measurement of $\tau_{\rm eff}$: bias-corrected measurement corrected for metal line absorption according to \cite{schaye_metallicity_2003} and optically thick absorbers according to \cite{becker_refined_2013}
    \item[d] Statistical uncertainty on $\tau_{\rm eff}$ estimated from our covariance matrix
    \item[e] Total statistical and systematic uncertainty (\S\ref{sec:taucorrs})
    \end{tablenotes}
    \end{threeparttable}
\end{table*}

\subsection{Comparison with the literature}\label{sec:taucomparisons}

Previous measurements of the optical depth have either used a relatively small number of high-resolution, high SNR spectra or a large number of more moderate resolution spectra with generally lower SNR. The high-resolution studies offer the promise of a direct measurement of the continuum between the absorption features, particularly at lower redshifts, while large samples of moderate resolution spectra offer greater statistical power. \cite{faucher-giguere_direct_2008} performed a direct measurement using 86 $z>2$ high-resolution quasar spectra. In contrast, \cite{becker_refined_2013} used 6,065 moderate resolution SDSS DR7 spectra for a relative measurement of the mean transmitted flux as a fraction of its value at $z \leq 2.5$ with the assumption that the mean unabsorbed quasar continuum does not evolve with redshift. They used results from \cite{faucher-giguere_direct_2008} to convert this to an absolute measurement. \cite{kamble_measurements_2020} used 40,035 SDSS DR12 quasar spectra to perform a similar relative measurement while also allowing for spectral diversity. They found a higher opacity at $z \leq 3$ compared to \cite{becker_refined_2013}. More recently, \cite{liu_deep_2021} used their feedforward neural network-based continuum predictions on $\sim3200$ SDSS DR16 quasar spectra to perform an absolute measurement of the effective optical depth. \cite{gaikwad_2021} used a sample of 103 high-resolution, high SNR quasar spectra from \textsc{kodiaq dr2} to directly measure the evolution of the mean flux, and found results in agreement with \cite{faucher-giguere_direct_2008}. Additionally, \cite{ding_new_2023} presented a new measurement of the effective Ly$\alpha$ forest opacity based on 27,008 SDSS DR14 quasar spectra using a Markov Chain Monte Carlo analysis. The authors found a systematic offset towards lower opacity values compared to \cite{kamble_measurements_2020}.

Some previous works reported evidence for a decrease in the effective optical depth at $z\simeq3.2$. Such a feature could be a signature of \ion{He}{2} reionization, as it could signal a rise in the IGM gas temperature at the end of helium reionization \citep[see e.g.][]{theuns_detection_2002,faucher-giguere_direct_2008,becker_refined_2013}. \cite{bernardi_feature_2003} found a sudden $\sim10$\% reduction in the effective optical depth at this redshift, which was later corroborated by \cite{faucher-giguere_direct_2008} despite substantial differences in their datasets and analysis methods. However, others have argued from a theoretical perspective that such a sharp feature should not be observed \citep[e.g.][]{bolton_evolution_2009}. Some temperature evolution measurements \citep[e.g.][]{becker_detection_2011,walther_2019} also suggest that IGM heating due to \ion{He}{2} reionization occurred over a more prolonged period, which would not produce a sharp feature in the evolution of the effective optical depth. However, more recent temperature measurements from \cite{gaikwad_2021} provide evidence for a more rapid \ion{He}{2} reionization. Additional measurements of the mean transmitted flux by \cite{becker_refined_2013} revealed no evidence for a departure from a smooth power-law description. Later works \citep[e.g.][]{kamble_measurements_2020,liu_deep_2021,ding_new_2023} also reported no evidence for this $z\simeq3.2$ feature.

\begin{figure*}
    \centering
    \includegraphics{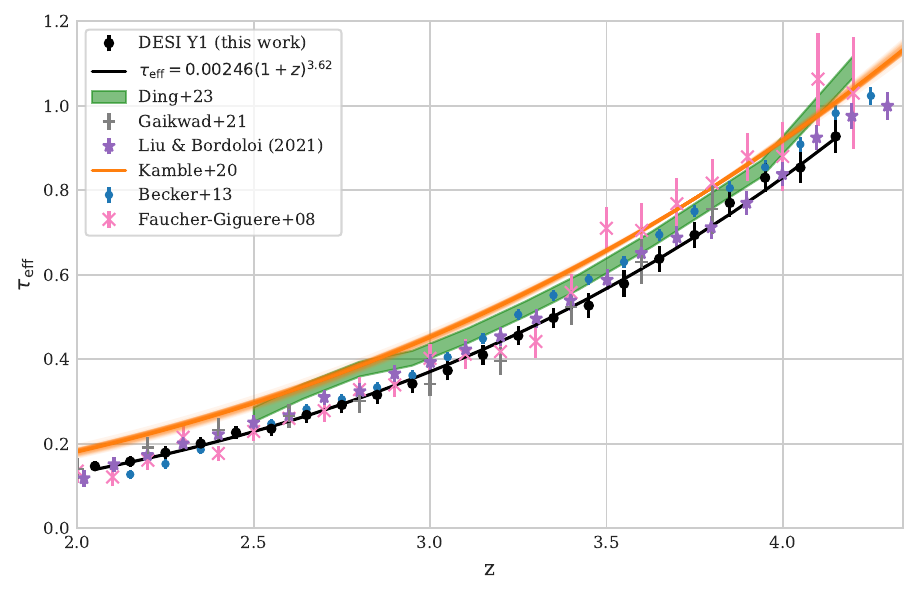} 
    \caption{Measurements of the effective optical depth and associated power-law (PL) fit with DESI Y1 data using LyCAN continuum predictions. We find the best agreement with measurements based on high-resolution spectra at lower redshifts ($z<3$) where continuum fitting errors are minimal \protect{\citep{faucher-giguere_direct_2008,gaikwad_2021}}. Our total uncertainty error bars are dominated by the systematic error due to the metal correction scheme, which is not reflected in some other results in the literature \protect{\citep[e.g.][]{becker_refined_2013,liu_deep_2021}} (see~\S\ref{sec:taucorrs}).}
    \label{fig:taueff}
\end{figure*}

Our results for the evolution of the mean optical depth show a smooth power-law relationship of the form $\tau_{\rm eff}(z) = \tau_0(1+z)^\gamma$ where $\tau_0 = (2.46 \pm 0.14)\times10^{-3}$ and $\gamma = 3.62 \pm 0.04$, and no evidence for a dip at $z\simeq3.2$. Our approach should be sensitive to this potential feature, as our systematic errors vary smoothly with redshift. We find the closest agreement with measurements based on high-resolution spectra \citep[e.g.][]{faucher-giguere_direct_2008,gaikwad_2021}, particularly at the lower redshift end. High-resolution spectra are subject to smaller errors in the continuum fitting process, and continuum errors are minimal at lower redshifts where Ly$\alpha$ forest absorption is reduced. This agreement indicates that LyCAN can accurately predict the underlying quasar continuum. We also measure a lower optical depth at higher redshifts ($z\gtrsim3.5$) compared to previous works, which may indicate a stronger ionizing background at these redshifts. Further, we find good agreement with \cite{liu_deep_2021}, whose measurements were also corrected for metals according to \cite{schaye_metallicity_2003}, over our whole redshift range. This points to the self-consistency of deep learning methods for quasar continuum prediction. Our measurements are shown alongside others in Figure~\ref{fig:taueff}. The error bars for the \cite{becker_refined_2013} and \cite{liu_deep_2021} measurements are likely underestimated as they do not include the systematic error in the correction for metal line absorption.

\section{Summary}\label{sec:summary}

We have developed LyCAN, a CNN-based approach to predict the underlying quasar continuum in the Ly$\alpha$ forest region strictly based on quasar spectra on the red side of the  Ly$\alpha$ emission line. We have shown that LyCAN performs extremely well on all testing data, and outperforms the PCA- and NMF-based predictive models developed for this work using the same training data. LyCAN also uses the flux on the red side of Ly$\alpha$ as input without any smoothing or additional preprocessing steps, while the PCA- and NMF-based models are more sensitive to noise and the presence of other features in quasar spectra, such as metal line absorption. 

We began our analysis with a sample of 48 archival COS observations. We generated NMF components for this sample and fit these components to a large sample of high-SNR (median $r-$band SNR $\geq$ 5) DESI Y1 spectra. The NMF coefficients for 10 of the COS spectra were very inconsistent with the coefficient distribution for the DESI quasars, and we therefore eliminated them from the sample. We used the remaining set of 38 low-redshift COS spectra to generate a sample of 40,000 unique synthetic spectra with varying levels of noise and emission line properties modeled after COS and DESI Y1 observations. We supplemented these with a sample of 40,000 DESI Y5 mock spectra, and randomly divided the resulting dataset of 80,000 spectra into training, testing, and validation subsets. 

We found the optimal CNN architecture through a pseudo-random search of hyperparameters. We optimized the number of convolutional layers, number of filters and kernel size per layer, as well as the number of fully-connected layers and number of nodes per layer. Our final architecture was chosen as the option that minimized both the median AFFE score (Eq.~\ref{affemscore}) and $1\sigma$ range score (Eq.~\ref{affepscore}). The LyCAN architecture features six hidden layers, including three convolutional layers and three dense layers. 

We used the LyCAN continuum predictions on DESI Y1 quasars to perform an absolute measurement of the evolution of the effective optical depth with the largest sample to date. Our results show close agreement with previous measurements based on high-resolution quasar spectra at lower redshifts \citep[e.g.][]{faucher-giguere_direct_2008,gaikwad_2021} where those methods are most able to measure the quasar continuum between the absorption features. This demonstrates that LyCAN can accurately predict the unabsorbed continuum. We find that the mean optical depth varies smoothly with redshift, and do not find evidence for a feature at $z\simeq3.2$ that could be associated with \ion{He}{2} reionization. Due to our large sample size of 83,635 quasars and LyCAN's ability to accurately predict the underlying continuum, we expect that this is the most accurate measurement of the effective optical depth evolution to date.

Our analysis is limited by the quality of our training set and its application to DESI quasars. The median AFFE on synthetic datasets is likely to be more of a lower bound on the uncertainty associated with real DESI data. More $UV$ spectra of high-luminosity, low-redshift quasars could substantially improve the empirical training set. Additionally, our correction procedure for the observed redshift bias in LyCAN predictions assumes that the DESI mock spectra are representative of the real Y1 data. This is a common assumption in large surveys such as DESI as mocks are used to quantify sources of systematic error. Future studies of spectral diversity in the quasar sample would help to better quantify this source of systematic error. Further, since LyCAN is a CNN, continuum predictions are deterministic and do not enable the quantification of uncertainties on real DESI data. Improvements on this, as well as direct comparison to other continuum prediction models in the literature, are left to future work.

LyCAN was primarily developed to be an alternative method to \texttt{picca} to measure the flux transmission field $\delta_q(\lambda)$ of the Ly$\alpha$ forest and measure the Ly$\alpha$ auto-correlation and Ly$\alpha$-QSO cross-correlation functions. We plan to use LyCAN continuum predictions on DESI Y1 quasars to compute the flux transmission field in the Ly$\alpha$ forest region in a future paper, and use these to compute cosmological parameters. This alternative approach will complement the standard analysis, especially as the measurements of $\delta_q(\lambda)$ with LyCAN will not have the same biases of the mean and spectral slope that distort the correlation function. This will help retain information on large scales for the one dimensional \citep{ravoux_dark_2023,karacayli_optimal_2024} and three dimensional \citep{karim_2023_p3d,roger_p3d_2024} power spectrum, and may be especially valuable for future work to measure the full shape of the auto-correlation function \citep{cuceu23a,cuceu23b}.

\begin{acknowledgments}
\section*{Acknowledgments}
WT and PM acknowledge support from the United States Department of Energy, Office of High Energy Physics under Award Number DE-SC-0011726. We thank Joe Antognini, David Weinberg, and Chris Hirata for helpful discussions, and Nicolas Lehner for sharing information about COS observations of quasars. 

This material is based upon work supported by the U.S. Department of Energy (DOE), Office of Science, Office of High-Energy Physics, under Contract No. DE–AC02–05CH11231, and by the National Energy Research Scientific Computing Center, a DOE Office of Science User Facility under the same contract. Additional support for DESI was provided by the U.S. National Science Foundation (NSF), Division of Astronomical Sciences under Contract No. AST-0950945 to the NSF’s National Optical-Infrared Astronomy Research Laboratory; the Science and Technology Facilities Council of the United Kingdom; the Gordon and Betty Moore Foundation; the Heising-Simons Foundation; the French Alternative Energies and Atomic Energy Commission (CEA); the National Council of Humanities, Science and Technology of Mexico (CONAHCYT); the Ministry of Science and Innovation of Spain (MICINN), and by the DESI Member Institutions: \url{https://www.desi.lbl.gov/collaborating-institutions}. Any opinions, findings, and conclusions or recommendations expressed in this material are those of the author(s) and do not necessarily reflect the views of the U. S. National Science Foundation, the U. S. Department of Energy, or any of the listed funding agencies.

The authors are honored to be permitted to conduct scientific research on Iolkam Du’ag (Kitt Peak), a mountain with particular significance to the Tohono O’odham Nation.
\end{acknowledgments}

\section*{Data Availability}
The data points for each figure are available at \dataset[doi:10.5281/zenodo.11106581]{https://doi.org/10.5281/zenodo.11106581}. The COS observations used for generation of synthetic spectra can be accessed via \dataset[doi: 10.17909/ae35-e630]{https://doi.org/10.17909/ae35-e630}.

\vspace{5mm}
\facilities{HST (COS), KPNO:Mayall (DESI).}
\software{Numpy \citep{harris_numpy_2020}, Scipy \citep{scipy_2020}, Matplotlib \citep{matplotlib_2007}, Keras \citep{keras_2018}, TensorFlow \citep{abadi2016tensorflow}, LineTools \citep{linetools_2017}, Astropy \citep{astropy_2013,astropy_2018,astropy_2022}, Scikit-Learn \citep{sklearn_2011}, NonnegMFPy \citep{zhu_nonnegative_2016}, Spectres \citep{spectres_2017}, Corner \citep{corner_2016}.}

\bibliography{references}{}
\bibliographystyle{aasjournal}

\end{document}